\documentclass{svmult}

\usepackage{makeidx}             
\usepackage{graphicx}             
\usepackage{multicol}               
\usepackage[bottom]{footmisc}

\makeindex             

\usepackage{amssymb,amsmath,psfig,amscd,xy}
\usepackage{mathrsfs}

\def\C{\mathbb C}

\def\E{\mathbb E}
\def\H{\mathbb H}
\def\I{\mathbb I}
\def\kk{\mathbb K}
\def\P{\mathbb P}
\def\Q{\mathbb Q}
\def\R{\mathbb R}
\def\S{\mathbb S}

\def\Z{\mathbb Z}

\def\Ga{\alpha}
\def\Gb{\beta}
\def\Ge{\varepsilon}
\def\Gg{\gamma}
\def\GG{\Gamma}
\def\Gd{\delta}
\def\GD{\Delta}
\def\GL{\Lambda}

\def\Gm{\mu}
\def\Go{\omega}
\def\GO{\Omega}
\def\Gr{\varrho}
\def\GP{\Phi}
\def\Gp{\phi}
\def\Gvp{\varphi}
\def\GS{\Sigma}
\def\Gs{\sigma}

\def\Gt{\tau}

\def\Bd{\mathbf d}
\def\BE{\mathbf E}
\def\BF{\mathbf F}

\def\Bj{\mathbf j}

\def\BS{\mathbf S}
\def\BT{\mathbf T}
\def\BV{\mathbf V}
\def\BY{\mathbf Y}
\def\BZ{\mathbf Z}

\def\BGe{\boldsymbol\varepsilon}
\def\BGg{\boldsymbol\gamma}

\def\BGe{\boldsymbol\eta}

\def\BGm{\boldsymbol\mu}

\def\BGt{\boldsymbol\tau}

\def\kBGg{\ff{\boldsymbol\gamma}}

\def\kBGm{\ff{\boldsymbol\mu}}

\def\kBGt{\ff{\boldsymbol\tau}}

\def\cC{\mathcal C}

\def\cG{\mathcal G}
\def\cH{\mathcal H}
\def\cI{\mathcal I}

\def\cO{\mathcal O}

\def\cR{\mathcal R}

\def\sD{\mathscr D}

\def\fD{\mathfrak D}

\def\fM{\mathfrak M}

\def\fR{\mathfrak R}

\def\dd{\partial}
\def\Bdd{\boldsymbol\dd}
\def\smin{\setminus}

\def\emp{\emptyset}
\def\eksi{\mathbf{id}}

\def\fix{\mathrm{Fix}}

\def\bfix{\mathbf{Fix}}
\def\bconj{\mathbf{Perm}}

\def\Nn{1,\cdots,n}
\def\nn{\underline{\mathbf n}}
\def\im{\sqrt{-1}}

\def\projc{\P^{1}(\C)}
\def\projr{\P^{1}(\R)}

\newtheorem{thmquote}{Theorem}

\def\tree{\mathcal Tree}

\newcommand{\ve}[1]{\mathbf{#1}}

\newcommand{\comp}[1]{#1(\C)}
\newcommand{\real}[1]{#1(\R)}

\def\curve{(\GS;\ve{p})}
\def\map{(\GS;\ve{p};f)}

\newcommand{\cmod}[1]{\overline{M}_{#1}(\C)}
\newcommand{\cmodo}[1]{M_{#1}(\C)}
\newcommand{\umod}[1]{\overline{U}_{#1}(\C) }
\newcommand{\cdiv}[1]{D_{#1}(\C)}
\newcommand{\cdivc}[1]{\overline{D}_{#1}(\C)}

\newcommand{\cmap}[1]{\overline{M}_{#1}(X,\Gb)}

\newcommand{\rmod}[1]{\overline{M}^{\Gs}_{#1}(\R)}

\newcommand{\rmodo}[1]{M^{\Gs}_{#1}(\R)}
\newcommand{\rdiv}[1]{D_{#1}(\R)}
\newcommand{\rdivc}[1]{\overline{D}_{#1}(\R)}

\newcommand{\rmap}[1]{\overline{R}^\Gs_{#1}(X,c_X,\Bd)}
\newcommand{\rmapo}[1]{R^\Gs_{#1}(X,c_X, \Bd)}

\newcommand{\csp}[1]{Conf_{(#1)}}
\newcommand{\cspp}[1]{{\widetilde{Conf}}_{(#1)}}

\newcommand{\cspq}[1]{\widetilde{C}_{(#1)}}
\newcommand{\csq}[1]{C_{(#1)}}
\newcommand{\csqc}[1]{\overline{C}_{#1}}

\newcommand{\konj}[1]{\overline{#1}}
\newcommand{\hh}[1]{\hat{#1}}

\newcommand{\ff}[1]{\widetilde{#1}}

\newcommand{\cls}[1]{[\csqc{#1}]}
\newcommand{\call}[1]{[ \overline{D}_{#1} ]}
\newcommand{\callr}[1]{[ \overline{D}_{#1}(\R) ]}

\begin{document}

\title*{Towards quantum cohomology of real varieties}

\author{\"Ozg\"ur Ceyhan}

\institute{Centre de Recherches Math\'ematiques, Universit\'e de Montr\'eal, Canada
\texttt{ceyhan@crm.umontreal.ca}}
%
%
\maketitle

\begin{abstract}
This paper is devoted to a discussion of Gromov-Witten-Welschinger (GWW) 
classes and their applications. In particular,  Horava's  definition of quantum 
cohomology  of real algebraic varieties is revisited by using GWW-classes and 
it is introduced as a DG-operad. In light of this definition, we speculate about 
mirror symmetry for real varieties.
\end{abstract}

\rightline{
\vbox{\hsize 70mm 
\noindent 
{\it The strangeness and absurdity of these replies arise from the 
fact that modern history, like a deaf man, answers questions 
no one has asked.}
\noindent
Leo Tolstoy, War and Peace.
 }}

\section{Introduction}

\subsection{Quantum cohomology of complex varieties} 
Let $X$ be a projective algebraic variety over $\C$, and let  $\cmod{\BS}$ be 
the moduli space of $\BS$-pointed (complex) stable curve of genus zero.

In their seminal work \cite{km1}, Kontsevich and Manin define the Gromov-Witten 
(GW) classes of the variety $X$  as a collection of linear maps
$$\{ I^X_{\BS,\Gb}:  \bigotimes_\BS H^*X \to H^* \overline{M}_{\BS} \}$$ that are 
expected to satisfy a series of formal and geometric properties. These 
invariants actually  give  appropriate enumerations of rational curves in $X$ 
satisfying certain incidence conditions.

The quantum cohomology of $X$ is a formal deformation of its cohomology ring. 
The parameters of this deformation are coordinates on the space $H^*(X)$, and 
the structure constants are the third derivatives of a formal power series $\GP$ 
whose  coefficients are the top-dimensional GW-classes of $X$.

Formal solution of $\GP$ of associativity, or WDVV, equation has provided a
source for solutions to problem in  complex enumerative geometry.

\bigskip
\noindent
{\it A refined algebraic picture: Homology operad of $\cmod{\BS}$.} 
By dualizing $\{I^X_{\BS,\Gb}\}$, a more refined  algebraic structure on $H^*X$
is obtained:
\begin{eqnarray*}
\BY_\BS: H_* \konj{M}_{\BS \cup \{s\}} \to Hom(\bigotimes_\BS H^*X, H^*X).
\end{eqnarray*}
In this picture, any homology class in $ \konj{M}_{\BS \cup \{s\}} $ is interpreted as an
$n$-ary operation on $H^*X$. The additive relations in $H_* \konj{M}_{\BS \cup \{s\}}$ 
become identities between these operations. Therefore,  $H^*X$ carries a structure of 
an algebra over the cyclic operad  $H_* \konj{M}_{\BS \cup \{s\}}$.

One of the remarkable basic results in the theory of the associativity equations 
(or Frobenius manifolds) is the fact that their formal solutions are the same as 
cyclic algebras over the homology operad $H_* \konj{M}_{\BS \cup \{s\}}$.

\subsection{Quantum cohomology of real varieties} 

Physicists have long been suspecting that there should be analogous algebraic 
structures in open-closed string theory arising from the enumeration of real rational 
curves/discs with Lagrangian boundary condition.  By contrast with the spectacular 
achievements of closed string theory in complex geometry, the effects of 
open-closed string theory in real algebraic geometry remained deficient
due to the lack of suitable enumerative invariants. The main obstacle with 
defining real enumerative invariants is that the
number of real objects usually varies along the parameter space. 

The perspective for the real situation  has 
radically changed after the discovery of Welschinger invariants. In a series 
paper \cite{w1}-\cite{w4}, J.Y. Welschinger introduced a set invariants for 
real varieties that give   lower bounds on the number of real solutions.

\bigskip
\noindent
{\it Welschinger invariants.} 
 Welschinger has defined a set of  invariants 
counting, with appropriate weight $\pm 1$, real rational $J$-holomorphic curves  
intersecting a generic real configuration (i.e., real or  conjugate pairs) of marked 
points. Unlike the usual homological definition of Gromov-Witten 
invariants, Welschinger invariants are originally defined  by assigning signs to 
individual curves based on certain geometric-topological criteria (such as
number of solitary double points in the case of surfaces, self-linking numbers in
the case of 3-folds etc). A homological interpretation of Welschinger invariants 
has been given by J. Solomon recently (see \cite{s}):

Let $(\comp{X},c_X)$ be a real variety and $\real{X}: = \fix(c_X)$ be its real part. 
Let $\rmap{\BS}$ be the moduli space of real stable maps which are invariant
under relabeling by the involution $\Gs$.

Let $\mu_*$  and $\Ga_*$  are Poincare duals 
of point classes (respectively in $\comp{X}$ and $\real{X}$).  Solomon
showed that Welschinger invariants $N^{\Gs}_{\BS,\Bd}$ can be defined 
in terms of the (co)homology of the moduli space of real maps $\rmap{\BS}$:
\begin{eqnarray}
\label{eqn_solomon}
N^{\Gs}_{\BS,\Bd}   := \int_{[\rmap{\BS}]} \left\{
\bigwedge_{\{s,\konj{s}\} \subset \bconj(\Gs)} ev_s^* (\mu_s)
\ \wedge \ 
\bigwedge_{s \in \bfix(\Gs)} ev_s^*(\Ga_s) \right\}
\end{eqnarray}
Here,  it is important to note that the moduli space $\rmap{\BS}$ has 
codimension one boundaries which is in fact one of the main difficulties
of the definition of open Gromov-Witten invariants. That's why the 
earlier studies on open Gromov-Witten invariants focus on homotopy invariants
instead of actual enumerative invariants (see, for instance \cite{fu}).

\bigskip
\noindent
{\it Quantum cohomology of real varieties.}  
The quantum cohomology for real algebraic varieties has been introduced 
surprisingly early, in 1993, by P. Horava in \cite{ho}. In his paper,
Horava describes a $\Z_2$-equivariant topological sigma model on a
real variety $(X,c_X)$ whose set of physical observables (closed and open
string states)   is  a direct sum 
of the cohomologies of $\comp{X}$ and $\real{X}$;
\begin{eqnarray*}
\cH_c \oplus \cH_o := H^*(\comp{X}) \oplus H^*(\real{X}).
\end{eqnarray*}
The analogue of the quantum cohomology ring in Horava's setting is a structure 
of $H^*(\comp{X})$-module structure on $H^*(\comp{X}) \oplus H^*(\real{X})$ 
obtained by deforming cup and mixed products.

Solomon's homological interpretation of Welschinger invariants opens a new
gate to reconsider the quantum cohomology of real varieties. In this paper, we define
the {\it Welschinger classes} by using Gromov-Witten theory as a guideline: 
By considering the following diagram
\begin{eqnarray*}
\begin{CD}
\rmap{\BS}  @> {ev} >>  \prod_{s \in \bconj(\Gs)/\Gs} \comp{X} 
					\times  \prod_{s \in \bfix(\Gs)} \real{X}  \\
@V{\nu}VV      \\
\rmod{\BS}
\end{CD}
\end{eqnarray*}
we give Welschinger classes as a family of linear maps
\begin{eqnarray*}
W^X_{\BS,\Bd}: 
\bigotimes_{\bconj(\Gs)/\Gs} H^*(\comp{X}) 
\bigotimes_{\bfix(\BS)} H^*(\real{X}; \det(T\real{X}))
\to 
H^*(\rmod{\BS},\fD)
\end{eqnarray*}
where $\rmod{\BS}$ is the moduli space pointed real stable curves and
$\fD$ is the union of its substrata of codimension one or higher.

In a recent paper \cite{c1}, we have shown that homology
of the moduli space $\rmod{\BS}$ is isomorphic to the homology of a combinatorial 
graph complex which is generated by the strata of $\rmod{\BS}$. By using a  reduced version of this graph complex $\cC_\bullet$, in this paper,
we introduce  {\it quantum cohomology of real variety} $(\comp{X},c_X)$. 
We construct a  differential graded (partial) operad
\begin{eqnarray*}
\{ \BZ_\BS^\Gs:  \cC_\bullet  \to 
Hom(\bigotimes_{\bconj(\Gs)/\Gs} \cH_c
\bigotimes_{\bfix(\Gs)} \cH_o, \cH_o) \}
\end{eqnarray*}
along with an algebra over $H_* \konj{M}_{\BS \cup \{s\}}$-operad 
\begin{eqnarray*}
\{ \BY_\BS:  H_* \konj{M}_{\BS \cup \{s\}} \to 
Hom(\bigotimes_{\BS} \cH_c, \cH_c) \}
\end{eqnarray*}
which serve as the {\it quantum cohomology of real variety} $(\comp{X},c_X)$.

\bigskip
\noindent
{\it A reconstruction theorem for Welschinger invariants.}  
Until very recently, the only calculation technique for Welschinger invariants was 
Mikhalkin's method which is based on tropical algebraic geometry.
(see \cite{mik1,mik2}).  By using Mikhalkin's technique, Itenberg, Kharlamov and 
Shustin proved a recursive formula for Welschinger invariants \cite{iks}. In their 
work, they also give a definition of higher genus Welschinger invariants in the tropical 
setting.

However, a real version of Kontsevich's recursive formula (for Welschinger
invariants) was still missing. In \cite{stalk}, Jake Solomon announced a 
differential equation satisfied by the generating function of Welschinger invariants and
give a recursion relation between Welschinger invariants of $\P^2(\C)$ with
standart real structure.

Solomon's method is very similar to the calculations of Gromov-Witten
invariants in complex situation i.e., considering certain degenerations and
homological relation between degenerate locus. However, the cycle
in the moduli of real maps which he considers, is quite intriguing. We 
recently noticed that the cycle, which leads to Solomon's formula, is a
combination of certain cycles  satisfying $A_\infty$ and Cardy relations.

The enumerative aspects of quantum cohomology of real varieties and 
its relation to Solomon's formula will be presented in a subsequent 
paper \cite{c2}.

\bigskip
\noindent
{\it `Realizing' mirror symmetry.}  
Kontsevich's conjecture of homological mirror symmetry and 
our definiton of quantum cohomology of real varieties as DG-operads, 
suggest a correspondence between two Ôopen-closed homotopy 
algebrasÕ (in an appropriate sense): 
\begin{center}
\begin{tabular}{c c c } 
\ \ \ \ \ Symplectic side $X$ \ (A-side) \ \ \ \ \ 
& $\Longleftrightarrow $   & 
\ \ \ \ \ Complex side $Y$ \ (B-side) \ \ \ \ \   \\
\hline \hline
\rule{0pt}{4ex}
$(\cH_c, \cH_o) =$
& $\Leftrightarrow $ & 
$(\widehat{\cH}_c, \widehat{\cH}_o) = $\\
\rule{0pt}{3ex}
$(H^{*}(X), HF^{*}(\real{X},\real{X}))$  
&  & 
$(H^{*,*}(Y), H^{0,*}(Y))$  \\
\end{tabular}
\end{center}

A brief account of the real version of mirror symmetry is given in the
final section of this paper and will be further discussed, especially its relation with
SYZ-picture, in a subsequent paper \cite{cmirror}.

\bigskip
\noindent
{\it Acknowledgements.} I take this opportunity to express my deep 
gratitude to Yuri Ivanovic Manin and Matilde Marcolli. They have been 
a mathematical inspiration for me ever since I met them, but as I've 
gotten to know them better, what have impressed me most are their 
characters. Their interest, encouragement and suggestions have been 
invaluable to me.

I also would like thank to D. Auroux and M. Abouzaid.  The last part
of this paper containing the discussion on mirror symmetry has been 
first inspired during a discussion with them at Gokova Conference.

I am thankful to  O. Cornea, C. Faber, F. Lalonde, A. Mellit, 
D. Radnell and A. Wand for their comments and suggestions.

This paper was conceived during my stay at Max-Planck-Institut f\"ur 
Mathematik (MPI), Bonn, its early draft was written at Centre de 
Recherhes Math\'ematiques (CRM), Montreal, and this final version 
was prepared during my stay at Mittag-Leffler Institute, Stockholm. Thanks 
are also due to these three institutes for their hospitality and support.



\subsubsection*{Table of contents:}

\begin{tabular}{ll}
{\bf 1}                    &{Introduction}  \\
{\bf Part I. }           &{\bf Moduli spaces of pointed complex and real curves} \\
{\bf 2}                    &{Moduli spaces of pointed complex curves of genus zero}  \\
{\bf 3}                    &{Moduli spaces of pointed real curves of genus zero}  \\
{\bf Part II. }           &{\bf Quantum cohomology of real varieties} \\
{\bf 4}                     &{Gromov-Witten classes}  \\
{\bf 5}                     &{Gromov-Witten-Welschinger classes}  \\
{\bf Part III. }           &{\bf Yet again, mirror symmetry} \\
\end{tabular}



\section*{Part I: Moduli spaces of pointed complex and real curves}

In order to define correlators of open-closed string theories, we 
need explicit descriptions of the homologies of both, the moduli 
space of pointed complex curves and moduli space  of pointed 
real curves. This section reviews the basic facts on pointed 
complex/real curves of genus zero, their moduli spaces and the 
homologies of these moduli spaces.

\subsubsection{Notation/Convention}
We denote the finite set $\{s_1,\cdots,s_n\}$ by $\BS$, and the
symmetric group consisting of all permutations of $\BS$ by $\S_n$.
We denote the set $\{\Nn\}$ simply by $\nn$.

In this paper, the genus of all curves is zero except when the
contrary is stated. Therefore, we usually  omit mentioning the
genus of curves.

\section{Moduli space of pointed complex curves}

An {\it $\BS$-pointed  curve} $\curve$ is a connected complex
algebraic curve $\GS$ with   distinct, smooth, {\it labeled points}
$\mathbf{p} = (p_{s_1},\cdots,p_{s_n}) \subset \GS$ satisfying the
following conditions:
\begin{itemize}
\item $\GS$ has only nodal singularities.%
\item The arithmetic genus of $\GS$ is equal to zero.
\end{itemize}
The nodal points and labeled points are called {\it special} points.

A {\it family of $\BS$-pointed  curves} over a complex
manifold $B(\C)$ is a proper, flat holomorphic map
$\pi_B: U_B(\C) \to B(\C)$ with
$n$ sections $p_{s_1},\cdots,p_{s_n}$ such that each geometric fiber
$(\GS(b);\mathbf{p}(b))$ is an $\BS$-pointed curve.

Two $\BS$-pointed curves  $\curve$ and $(\GS';\mathbf{p}')$ are {\it
isomorphic} if there exists a bi-holomorphic equivalence $\Phi: \GS
\to \GS'$ mapping $p_{s}$ to $p'_{s}$ for all $s \in \BS$.

An $\BS$-pointed curve is {\it stable} if its automorphism group is
trivial (i.e., on each irreducible component, the number of singular
points plus the number of labeled points is at least three).

\subsubsection{Graphs}

A {\it graph} $\Gg$ is a pair of finite sets of {\it vertices}
$\BV_{\Gg}$ and {\it flags} (or {\it half edges}) $\BF_{\Gg}$ with a
boundary map $\Bdd_{\Gg}: \BF_{\Gg} \to \BV_{\Gg}$ and an involution
$\Bj_{\Gg}: \BF_{\Gg} \to \BF_{\Gg}$ ($\Bj_{\Gg}^{2} =\eksi$). We
call $\BE_{\Gg}= \{(f_{1},f_{2}) \in \BF_{\Gg}^{2} \mid f_{1}=
\Bj_{\Gg}(f_{2})\ \& \ f_{1} \ne f_{2}\}$ the set of {\it edges},
and $\BT_{\Gg} = \{f \in \BF_{\Gg} \mid f = \Bj_{\Gg}(f)\}$ the set
of {\it tails}. For a vertex $v \in \BV_{\Gg}$, let $\BF_{\Gg}(v)
=\Bdd^{-1}_{\Gg}(v)$ and $|v| = |\BF_{\Gg}(v)|$ be the {\it valency}
of $v$.

We think of a graph $\Gg$ in terms of its {\it geometric
realization} $||\Gg||$ as follows: Consider the disjoint union of closed
intervals $\bigsqcup_{f_i \in \BF_{\Gg}} [0,1] \times f_i$, and
identify $(0, f_i)$ with $(0, f_j)$ if $\Bdd_\Gg(f_i) =
\Bdd_\Gg(f_j)$, and identify $(t, f_i)$ with $(1-t, \Bj_\Gg(f_i))$
for $t \in [0,1]$ and $f_i \ne \Bj_\Gg(f_i)$. The geometric realization of
$\Gg$ has a piecewise linear structure.

A {\it tree} is a graph whose geometric realization
is connected and simply-connected. If $|v|>2$ for all vertices, then
such a tree is called {\it stable}.

There are only finitely many isomorphism classes of stable trees
whose set of tails $\BT_{\Gg} $ is equal to $\BS$ (see, \cite{m} or \cite{c}).
We call the isomorphism classes of such trees {\it $\BS$-trees}.

\subsubsection{Dual trees of $\BS$-pointed curves}

Let $\curve$ be an $\BS$-pointed stable curve and $\eta: \hat{\GS}
\to \GS$ be its normalization. Let
$(\hat{\GS}_v;\hat{\mathbf{p}}_v)$ be the following pointed stable
curve: $\hat{\GS}_{v}$ is a component of $\hat{\GS}$, and
$\hat{\mathbf{p}}_v$ is the set of points consisting of the
pre-images of special points on $\GS_{v} := \eta (\hat{\GS}_v)$. The
points $\hat{\ve{p}}_v = (p_{f_{1}},\cdots, p_{f_{|v|}})$ on
$\hat{\GS}_v$ are ordered by the elements $f_*$ in the set
$\{f_{1},\cdots, f_{|v|}\}$.

The {\it dual tree} of an $\BS$-pointed stable
curve $\curve$ is an $\BS$-tree $\Gg$   where
\begin{itemize}
\item $\BV_{\Gg}$  is the set of components of $\hat{\GS}$.

\item $\BF_{\Gg}(v)$ is the set consisting of the pre-images of special
points in $\GS_v$.

\item $\Bdd_{\Gg}: f \mapsto v$ if and only if $p_f \in \hat{\GS}_v$.

\item $\Bj_{\Gg}: f \mapsto f$ if and only if $\eta(p_f)$ is a labeled
point, and $\Bj_{\Gg}: f_1 \mapsto f_2$ if and only if $p_{f_1} \in
\hat{\GS}_{v_1}$ and $p_{f_2} \in  \hat{\GS}_{v_2}$ are
the pre-images of a nodal point $\GS_{v_1} \cap \GS_{v_2}$.
\end{itemize}

\subsubsection{Combinatorics of degenerations: Contraction morphisms of $\BS$-trees}

Let $\curve$ be an $\BS$-pointed stable curve whose  dual tree is
$\Gg$. Consider the deformations of a nodal point of  $\curve$. Such
a deformation of $\curve$ gives a {\it contraction} of an edge of
$\Gg$: Let $e=(f_{e},f^{e}) \in \BE_{\Gg}$ be the edge corresponding
to the nodal point which is deformed, and let  $\Bdd_{\Gg}(f_e) = v_{e},
\Bdd_{\Gg}(f^e) =v^{e}$.
Consider the equivalence relation $\thicksim$ on the set of
vertices, defined by: $v \thicksim v$ for all $v \in \BV_{\Gg} \smin
\{v_{e},v^{e}\}$, and  $v_{e} \thicksim v^{e}$. Then there is an
$\BS$-tree $\Gg/e$ whose set of  vertices is $\BV_{\Gg}/\thicksim$ and
whose set of flags is $\BF_{\Gg} \smin \{f_e,f^e\}$.  The boundary map and
involution of $\Gg/e$ are the restrictions of $\Bdd_{\Gg}$ and
$\Bj_{\Gg}$.

We use the notation $\Gg < \Gt$ in order to indicate that the
$\BS$-tree $\Gt$ is obtained by contracting a set of edges of $\Gg$.

\subsection{Moduli space of $\BS$-pointed curves}
\label{sec_c_moduli}

The moduli space $\cmod{\BS}$ is the space of  isomorphism classes
of $\BS$-pointed stable curves. This space is stratified according
to the degeneration types of $\BS$-pointed stable curves. The degeneration types
of $\BS$-pointed stable curves are combinatorially encoded by $\BS$-trees.
In particular, the principal stratum $\cmodo{\BS}$ parameterizes
$\BS$-pointed irreducible curves; i.e., it is associated to the one-vertex
$\BS$-tree. The principal stratum $\cmodo{\BS}$  is the quotient of the product
$(\projc)^{n}$ minus the diagonals $\GD = \bigcup_{k<l} \{(p_{s_1},
\cdots,p_{s_n})| p_{s_k} = p_{s_l} \}$ by $Aut(\projc) = PSL_2(\C)$.

\begin{thmquote}[Knudsen \& Keel, \cite{knu,ke}]
\label{thm_complex_moduli}

(a) For any $|\BS| \geq 3$,  $\cmod{\BS}$ is a smooth projective
algebraic variety of (real) dimension $2|\BS|-6$.

(b)  Any family of $\BS$-pointed stable curves over $B(\C)$ is induced
by a unique morphism $\kappa: B(\C) \to \cmod{\BS}$. The universal
family of curves $\umod{\BS}$ of $\cmod{\BS}$ is isomorphic to
$\cmod{\BS \cup \{s_{n+1}\}}$.

(c) For any $\BS$-tree $\Gg$, there exists a quasi-projective
subvariety  $\cdiv{\Gg} \subset \cmod{\BS}$ parameterizing the $\BS$-pointed
curves whose dual tree is given by $\Gg$. The subvariety $\cdiv{\Gg}$ is isomorphic to
$\prod_{v \in \BV_{\Gg}} \cmodo{\BF_\Gg(v)}$. Its (real) codimension
(in $\cmod{\BS}$) is $2|\BE_{\Gg}|$.

(d) $\cmod{\BS}$ is stratified by pairwise disjoint subvarieties
$\cdiv{\Gg}$. The closure $\cdivc{\Gg}$ of any stratum $\cdiv{\Gg}$  is stratified by
$\{\cdiv{\Gg'} \mid \Gg' \leq \Gg \}$.
\end{thmquote}

\subsubsection{Forgetful morphisms}
\label{sec_forgetful}

We say that $(\GS;p_{s_1},\cdots,p_{s_{n-1}})$ is obtained by
forgetting the labeled point $p_{s_n}$ of an $\BS$-pointed curve
$(\GS;p_{s_1},\cdots, p_{s_n})$. However, the resulting pointed
curve may well be unstable. This happens when the component
$\GS_{v}$ of $\GS$ supporting $p_{s_n}$ has only two additional
special points. In this case, we contract this component to its
intersection point(s) with the components adjacent to $\GS_{v}$.
With this {\it stabilization}, we extend this map to the whole space and
obtain $\pi_{\{s_n\}}: \cmod{\BS} \to \cmod{\BS'}$ where $\BS'= \BS
\smin \{s_n\}$.  There exists a canonical isomorphism $\cmod{\BS}
\to \umod{\BS'}$ commuting with the projections to $\cmod{\BS'}$. In
other words, $\pi_{\{s_n\}}: \cmod{\BS} \to \cmod{\BS'}$ can be
identified with the universal family of curves.

A very detailed study on the moduli space $\cmod{\BS}$ can be found
in Chapter 3.2 and 3.3 in \cite{m}, and also in \cite{ke,knu}.

\subsection{Intersection ring of $\cmod{\BS}$} 
\label{homology}

In \cite{ke}, Keel gave a construction of the moduli space  $\cmod{\BS}$  
via a sequence of blowups of $\cmod{\BS \smin \{s_n\}} \times \projc$ 
along the certain (complex)  codimension two  subvarieties. This inductive 
construction of $\cmod{\BS}$ allowed him to calculate the intersection 
ring in terms of the divisor classes $[\cdivc{\Gg}]$. Note that 
the divisors $\cdiv{\Gg}$ parameterize $\BS$-pointed 
curves whose dual trees have only one edge.

For $|\BS| \geq 4$, choose $i,j,k,l \in \BS$, and let $\Gg,\Gt \in \tree$ such 
that $\Gt \not\approx \Gg$ and $|\BE_{\Gt}|=|\BE_{\Gg}|=1$. We write 
$ij \Gg kl$ if tails labeled by $i,j$ and $k,l$ belongs to different vertices of
$\Gg$. We call $\Gg$ and $\Gt$ {\it compatible} if there is no $\{i,j,k,l\} 
\subset \BS$ such that simultaneously $ij \Gg kl$ and $ik \Gt jl$.

From now on, we denote the divisor classes $[\cdivc{\Gg}]$ simply by $\call{\Gg}$.

\begin{thmquote}[Keel, \cite{ke}] 
\label{thm-ring}
For $|\BS| \geq 3$,
\begin{equation}
H_{*} (\cmod{\BS}; \Z) = \Z[ \ \call{\Gg}  \mid \Gg \in \tree,\ |\BE_{\Gg}|=1] / I_{\BS}
\nonumber
\end{equation}
is a graded polynomial ring, $deg\ \call{\Gg}=1$. The ideal $I_{\BS}$ is generated by the following relations:
\begin{enumerate}
\item For any distinct four elements $i,j,k,l \in \BS$:
\begin{equation}
\sum_{ij \Gg kl} \call{\Gg} -
\sum_{ik \Gt jl} \call{\Gt} = 0. \nonumber
\end{equation}
\item $\call{\Gg} \cdot \call{\Gt} = 0$ unless $\Gg$ and
$\Gt$ are compatible.
\end{enumerate}
\end{thmquote}

\subsubsection{Additive and multiplicative structures of $H_{*} (\cmod{\BS})$} 
\label{sec_intersection}

The precise description of homogeneous elements in $H_{*} (\cmod{\BS},\Z)$ is 
given by Kontsevich and Manin in \cite{km2}. The monomial $\call{\Gg_1} \cdots 
\call{\Gg_d}$ is called {\it good}, if $|\BE_{\Gg_i}|=1$ for all $i$, and $\Gg_i$'s 
are pairwise compatible. Consider any $\BS$-tree $\Gg$. Any edge $e \in \BE_{\Gg}$ 
defines an $\BS$-tree  $\Gg(e)$ which is obtained by contracting all edges of 
$\Gg$ but $e$. Then, we can associate a good  monomial 
\begin{equation*}
\call{\Gg} := \prod_{e \in \BE_{\Gg}} \call{\Gg(e)}
\end{equation*} 
of degree $|\BE_{\Gg}|$ to $\Gg$. The map $\Gg \mapsto \call{\Gg}$ establishes a 
bijection between the good monomials of degree $d$ in $H_{*}(\cmod{\BS},\Z)$, 
and $\BS$-trees  $\Gg$ with $|\BE_{\Gg}|=d$ (see \cite{km2}). Since boundary divisors 
intersect transversally when their trees are pairwise compatible, the good monomials 
are represented by the corresponding closed strata.

\begin{thmquote}[Kontsevich and Manin, \cite{km2}]
The classes of good monomials linearly generate $H_{*}(\cmod{\BS};\Z)$.
\end{thmquote}

\paragraph {Multiplication in $H_*(\cmod{\BS})$}
Let $\Gt,\Gg \in \tree$ and $|\BE_{\Gt}|=1$. In \cite{km2}, a product  formula 
of $\call{\Gt} \cdot \call{\Gg}$ is given in three distinguished cases:

\begin{itemize}
\item[A] Suppose that there exists an $e \in \BE_{\Gg}$ such that $\Gg(e)$ and 
$\Gt$ are not compatible (i.e., $\cdivc{\Gt} \cap \cdivc{\Gg(e)} = \emp$).
 Then $\call{\Gt} \cdot \call{\Gg}=0$.

\item[B] Suppose that $\call{\Gt} \cdot \call{\Gg}$ is a good monomial i.e., $\Gt, \Gg(e)$'s 
are pairwise compatible for all $e \in \BE_{\Gg}$. Then there exists a unique 
$\BS$-tree $\Gt'$ with $e' \in \BE_{\Gt'}$  such that $\Gt'/ e' = \Gg$, $\Gt'(e)=\Gt$, 
and $\call{\Gt} \cdot \call{\Gg} = \call{\Gt'}$.

\item[C] Suppose now that  there exists an $e \in \BE_{\Gg}$ such that $\Gg(e) = \Gt$ 
i.e., $\call{\Gt}$ divides $\call{\Gg}$.  For a given quadruple $\{i,j,k,l\}$ such that $ij  \Gt kl$, 
 we have
\begin{eqnarray} 
 \sum_{ij\Gt_1 kl} \call{\Gt_1} \cdot \call{\Gg} - \sum_{ik \Gt_2 jl} \call{\Gt_2} \cdot \call{\Gg}
 = 0. \nonumber
\end{eqnarray}
Since the elements of the second sum are not compatible with $D^{\Gg}$,
we have
\begin{equation}
\call{\Gt} \cdot \call{\Gg} = - \sum_{ {\Gt_1 \not= \Gt} 
\atop {ij \Gt_1  kl} }    \call{\Gt_1} \cdot \call{\Gg}.
\nonumber
\end{equation}
Here, $\call{\Gt_1} \cdot \call{\Gg}$ are good monomials, so they can be computed  as in 
previous case (B).
\end{itemize}

\paragraph {Additive relations of $H_{*}(\cmod{\BS})$}
\label{sec_additive}

It remains to give the linear relations between degree $d$ monomials. In \cite{km2}, 
these relations are given  in the following way. Consider an $\BS$-tree $\Gg$ with 
$|\BE_{\Gg}| = d-1$, and a vertex $v \in \BV_{\Gg}$ with $|v| \geq 4$. Let 
$f_{1}, f_{2}, f_{3}, f_{4} \in \BF_{\Gg} (v)$ be pairwise distinct flags. Put 
$\BF = \BF_{\Gg}(v) \setminus \{f_{1}, f_{2}, f_{3}, f_{4} \}$ and let $\BF_1,\BF_{2}$ 
be two disjoint subsets of $\BF$ such that $\BF = \BF_{1} \bigcup \BF_{2}$.  
We define two $\BS$-trees $\Gg_1, \Gg_2$.  The $\BS$-tree $\Gg_1$ is obtained 
by inserting a new edge $e=(f_e,f^e)$ to $\Gg$ at $v$  with boundary 
$\Bdd_{\Gg_1} (e) = \{v_e,v^e\}$ and flags  
$\BF_{\Gg_1}(v_e) =  \BF_{1} \bigcup \{f_{1},f_{2},f_e\}$ and 
$\BF_{\Gg_1}(v^e) =  \BF_{2} \bigcup \{f_{3},f_{4},f^e\}$.  The 
$\BS$-tree $\Gg_2$ is also obtained by inserting an edge $e$ to $\Gg$ at the same 
vertex $v$, but the flags are distributed differently on vertices 
$\Bdd_{\Gg_2}(e)=\{v_e,v^e\}$: 
$\BF_{\Gg_2}(v_e) =  \BF_{1}\bigcup \{f_{1},f_{3},f_e\}$ and
$\BF_{\Gg_2}(v^e) =  \BF_{2}\bigcup \{f_{2},f_{4},f^e\}$. Put 
\begin{equation}
\label{eqn_relations}
R(\Gg,v;f_{1},f_{2},f_{3},f_{4}) :=
\sum_{\Gg_{1}} \call{\Gg_{1}} - \sum_{\Gg_{2}} \call{\Gg_{2}}
\end{equation}
where summation is taken over all possible $\Gg_1$ and $\Gg_2$  
for fixed set of flags $\{f_1,\cdots,f_4\}$ as given above.

\begin{thmquote}[Kontsevich and Manin, \cite{km2}]
All linear relations between good monomials of degree $d$ are spanned by 
$R(\Gg,v;f_{1},f_{2},f_{3},f_{4})$  with  $|\BE_{\Gg}|=d-1$.
\end{thmquote}

A very detailed study on the moduli space $\cmod{\BS}$ can be found
in Chapter 3.2 and 3.3 in \cite{m}, and also in \cite{ke,knu}.

\section{Moduli space of pointed real curves}

In this section, we review some basic facts on the moduli spaces 
of $\BS$-pointed real curves  of genus zero  and their homology groups.

\subsection{Real structures of complex varieties}
A {\it real structure} on a complex variety $\comp{X}$ is an anti-holomorphic 
involution $c_X: \comp{X} \to \comp{X}$.  The fixed point set $\real{X}:= \fix(c_X)$
of the involution $c_X$ is called the {\it real part} of the variety $\comp{X}$
(or of the real structure $c_X$).

\subsubsection{Notation/Convention}
For an involution $\Gs \in \S_n$, we denote the subset $\{s \in \BS
\mid s=\Gs(s)\}$ by $\bfix(\Gs)$ and its complement $\BS \smin \bfix(\Gs)$
by  $\bconj(\Gs)$.

From now on, we assume that the involution $\Gs$ has fixed points i.e.,  
$\bfix(\Gs) \ne \emp$.\footnote{Welschinger invariants for $\Z_2$-equivariant 
point configurations with $\bfix(\Gs) = \emp$ are known to be zero (see \cite{w1,w2}). 
Moreover, the homology of the moduli space for these cases requires slightly different
presentation. Therefore, we exclude these special cases.}

\subsection{$\Gs$-invariant curves and their families}
\label{sec_s_curve}

An $\BS$-pointed stable  curve $\curve$ is called {\it
$\Gs$-invariant}  if it admits a real structure $c_\GS: \GS \to \GS$
such that $c_\GS(p_s) =p_{\Gs(s)}$ for all $s \in \BS$.

Let $\pi_B: \comp{U_B} \to \comp{B}$ be a  family of $\BS$-pointed
stable curves with a pair of real structures
\begin{eqnarray*}
\begin{CD}
\comp{U_B}  @> {c_{U}} >>   \comp{U_B}      \\
@V{\pi_B}VV      @VV{\pi_B}V  \\
\comp{B}       @> {c_B}   >>    \comp{B}.
\end{CD}
\end{eqnarray*}
Such a family is called  {\it $\Gs$-equivariant} if the following conditions
are met;
\begin{itemize}
\item if $\pi^{-1}(b) = \GS$, then $ \pi^{-1}(c_B(b)) = \konj{\GS}$
for every $b \in B$;

\item $c_U:  z \in \GS = \pi^{-1}(b) \mapsto z \in \konj{\GS}= \pi^{-1}(c_B(b))$.
\end{itemize}
Here a complex curve $\GS$ is regarded as a pair
$\GS=(C,J)$, where $C$ is the underlying two-dimensional manifold, $J$ is a complex structure on $C$, and  $\overline{\GS} = (C,-J)$
is its complex conjugated pair.

\begin{remark}
If $\pi_B: \comp{U_B} \to \comp{B}$ is a $\Gs$-equivariant family, then
each $(\GS(b),\ve{p}(b))$ for $b \in \real{B}$ is $\Gs$-invariant. It follows from the fact
that the group of automorphisms of $\BS$-pointed stable curves is
trivial.
\end{remark}

\begin{remark}
Since we have set $\bfix(\Gs) \ne \emp$, all $\Gs$-invariant curves 
are of type 1, i.e., their real parts  $\real{\GS}$ are trees of real projective 
spaces having   nodal singularities. This follows from the fact that real 
parts of $\Gs$-invariant curves cannot be the empty set (i.e., they can't be
type 2) and all special points must be distinct (i.e., they cen not have real
isolated nodal points either).

By contrast, $\Gs$-invariant curves can be of both type 1, type 2 or 
can have real isolated node when $\bfix(\Gs) = \emp$ (see \cite{c1}).
\end{remark}

\subsubsection{Combinatorial types of $\Gs$-invariant curves}

$\Gs$-invariant curves inherit additional structures on their sets of
special points. In this subsection, we first introduce the  `oriented' versions
of these structures.

Let $\curve$ be a $\Gs$-invariant curve, and let   $\Gg$ be its dual tree.
We denote the set of real components $\{v \mid c_\GS (\GS_v) = \GS_v\}$
of $\curve$ by $\BV_\Gg^\R$.

\paragraph{Oriented combinatorial types}

Let $(\hh{\GS};\hh{\ve{p}})$ be the normalization of a
$\Gs$-invariant curve $\curve$. By identifying
$\hh{\GS}_v$ with  $\GS_v \subset \GS$,  we obtain a real structure
on $\hh{\GS}_v$ for a real component $\GS_v$. The real part
$\real{\hh{\GS}_v}$ of this real structure divides $\hh{\GS}_v$ into
two halves: two 2-dimensional open discs, $\GS^+_v$ and $\GS^-_v$,
having $\real{\hh{\GS}_v}$ as their common boundary in $\hh{\GS}_v$.
The real structure $c_\GS: \hh{\GS}_v \to \hh{\GS}_v$ interchanges
$\GS^\pm_v$, and  the complex orientations of $\GS^\pm_v$ induce two
opposite orientations on $\real{\hh{\GS}_v}$, called its {\it complex
orientations}.

If we fix a labeling of halves of $ \hh{\GS}_v$ by $\GS^\pm_v$ (or
equivalently,  if we orient   $\real{\hh{\GS}_v}$ with one of the
complex orientations),  then the set of pre-images of special
points $\hh{\ve{p}}_v \in  \hh{\GS}_v$
admits the following structures:

\begin{itemize}
\item {\it An oriented cyclic ordering on the set of special points
lying in $\real{\GS_v}$:} For any point $p_{f_i} \in (\hat{\ve{p}}_v
\cap \real{\hh{\GS}_v})$, there is a unique $p_{f_{i+1}} \in
(\hat{\ve{p}}_v \cap \real{\hh{\GS}_v})$ which follows the point
$p_{f_i}$ in the positive direction of $\real{\hh{\GS}_v}$ (the
direction which is determined by the complex orientation induced
by the orientation of $\GS^+_v$).

\item {\it An ordered two-partition of the set of special points
lying in $\GS_v \smin \real{\GS_v}$}. The subsets $\hat{\ve{p}}_v
\cap \GS_v^\pm$ of $\hat{\ve{p}}_v$ give an ordered  partition
of $\hat{\ve{p}}_v \cap (\GS_v \smin \real{\GS_v})$ into two disjoint
subsets.
\end{itemize}

The relative positions of the special points lying in $\real{\hh{\GS}_v}$
and the complex orientation of  $\real{\hh{\GS}_v}$  give an
{\it oriented cyclic ordering} on the corresponding labeling set
$\BF_\Gg^\R(v) := (\hat{\ve{p}}_v \cap \real{\hh{\GS}_v})$. 
Moreover, the partition  $\{p_f \in \GS_v^\pm\}$ gives an {\it ordered two-partition}
$\BF_\Gg^\pm(v) := \{f \mid p_f \in \GS_v^\pm\}$ of
$\BF_\Gg (v) \smin \BF_\Gg^\R(v)$.

The {\it oriented combinatorial type} of a real component $\GS_v$ with a
fixed complex orientation is the following set of data:
\begin{eqnarray*}
o_v := \{\mathrm{two\ partition}\ \BF_\Gg^\pm(v);
\mathrm{oriented\ cyclic\ ordering\ on}\ \BF_\Gg^\R(v)\}.
\end{eqnarray*}

If we consider a $\Gs$-invariant curve $\curve$ with a  fixed complex
orientation at each real component, then the set of oriented combinatorial
types of real components
\begin{eqnarray*}
o := \{o_v \mid v \in \BV_\Gg^\R  \}
\end{eqnarray*}
is called an {\it oriented combinatorial type} of $\curve$.

\paragraph{Un-oriented combinatorial types}

The definition of oriented combinatorial types requires additional
choices (such as complex orientations) which are not determined by real structures of
$\Gs$-invariant curves.  By identifying the oriented combinatorial types for
such different choices, we obtain {\it unoriented combinatorial types}
of $\Gs$-invariant curves.
Namely, for each real component $\GS_v$ of a $\Gs$-invariant curve $\curve$, 
there are two possible ways of choosing $\GS_v^+$ in
$\hh{\GS}_v$. These two different choices give the {\it opposite}
oriented combinatorial types $o_v$ and $\bar{o}_v$: The
oriented combinatorial type $\bar{o}_v$ is obtained from $o_v$ by
reversing the cyclic ordering of $\BF_\Gg^\R(v)$ and swapping
$\BF_\Gg^+(v)$ and $\BF_\Gg^-(v)$.

The {\it unoriented combinatorial type} of a real
component $\GS_v$ of $\curve$  is the pair of opposite oriented
combinatorial types $u_v := \{o_v,\bar{o}_v\}$. The set of unoriented
combinatorial  types of real components
\begin{eqnarray*}
u := \{u_v \mid v \in \BV_\Gg^\R  \}
\end{eqnarray*}
is called the {\it unoriented combinatorial type} of $\curve$ .

\subsubsection{Dual trees  of $\Gs$-invariant curves}

Let $\curve$ be an $\Gs$-invariant curve and let $\Gg$ be its dual
tree.

\paragraph{O-planar trees}

An {\it oriented locally planar (o-planar) structure} on $\Gg$
is a set of data which encodes an oriented combinatorial type of
$\curve$. O-planar structures are explicitly given as follows:

\begin{itemize}
\item $\BV_{\Gg}^\R \subset \BV_\Gg$ is the set of real components
of $\GS$ (i.e., the set of {\it real vertices}).

\item $\BF_{\Gg}^\R (v) \subset \BF_\Gg(v)$ is the set of the
pre-images of special points in $\real{\GS_v}$. (i.e., the set of
{\it real flags} adjacent to the real vertex $v \in \BV_{\Gg}^\R$).

\item $\BF_{\Gg}^\R(v)$ carries an oriented cyclic ordering  for every
$v \in \BV_{\Gg}^\R$.

\item $\BF_\Gg(v) \smin \BF_{\Gg}^\R(v)$ carries an ordered two-partition
for every $v \in \BV_{\Gg}^\R$.
\end{itemize}

We denote $\BS$-trees $\Gg,\Gt,\Gm$ with o-planar structures
by $(\Gg,o),(\Gt,o),(\Gm,o)$ or
by bold Greek characters with tilde
$\kBGg,\kBGt,\kBGm$. When it is necessary to indicate different
o-planar structures on the same $\BS$-tree, we use indices in parentheses (e.g., $\kBGt_{(i)}$).

\paragraph{Notations.}

For each vertex  $v \in \BV_{\Gg}^\R$ (resp. $v \in \BV_{\Gg} \smin
\BV_{\Gg}^\R$) of an o-planar tree $\kBGg$, we associate a subtree
$\kBGg_{v}$ (resp. $\Gg_v$) which is given by $\BV_{\Gg_{v}} =
\{v\}, \BF_{\Gg_{v}} = \BF_{\Gg}(v)$, $\Bj_{\Gg_{v}} = \eksi$,
$\Bdd_{\Gg_{v}} = \Bdd_\Gg$, and by the  o-planar structure $o_v$ of
$\kBGg$ at the vertex $v \in \BV_\Gg^\R$.

A pair of vertices $v,\bar{v} \in \BV_\Gg \smin \BV_\Gg^\R$ is said to
be {\it conjugate} if $c_\GS(\GS_v) = \GS_{\bar{v}}$. Similarly, we
call a pair of flags $f,\bar{f} \in \BF_\Gg \smin \BF_\Gg^\R$ {\it
conjugate} if $c_\GS$ swaps the corresponding special points.

To each o-planar tree $\kBGg$,
we associate the subsets of vertices $\BV_\Gg^\pm$ and flags
$\BF_\Gg^\pm$  as follows:
Let $v_1 \in \BV_\Gg \smin \BV_\Gg^\R$, and let $v_2 \in \BV_\Gg^\R$
be the closest vertex to $v_1$ in $||\Gg||$. Let $f(v_1) \in
\BF_\Gg(v_2)$ be in the shortest path  connecting the vertices $v_1$
and $v_2$. The sets $\BV_\Gg^\pm$ are the subsets of vertices $v_1 \in
\BV_\Gg \smin \BV_\Gg^\R$ such that the corresponding flags $f(v_1)$ are
respectively in $\BF_\Gg^\pm(v_2)$. The subsets of
flags $\BF_\Gg^\pm$ are defined as
$\Bdd_\Gg^{-1}(\BV_\Gg^\pm)$.

\paragraph{U-planar trees}

A u-planar structure on the dual tree $\Gg$ of $\curve$ is the set of
data encoding the unoriented combinatorial type of $\curve$. It is
given by
\begin{eqnarray*}
u := \{(\Gg_v,o_v), (\Gg_v,\bar{o}_v) \mid v \in \BV_\Gg^\R\} 
\end{eqnarray*}

We denote $\BS$-trees  $\Gg,\Gt,\Gm$ with u-planar structures by
$(\Gg,u),(\Gt,u),(\Gm,u)$ or simply by  bold Greek characters
$\BGg,\BGt,\BGm$. O-planar planar trees $\kBGg,\kBGt,\kBGm$
give representatives of u-planar trees $\BGg,\BGt,\BGm$ respectively.

\subsubsection{Contraction morphism of o/u-planar trees}

\paragraph{Contraction morphism of o-planar trees}
Consider a family of $\Gs$-invariant curves
which is a deformation of a real node of the central fiber
$(\GS(b_0),\ve{p}(b_0))$ with a given oriented combinatorial type.
Let $\kBGt,\kBGg$ be the o-planar trees associated respectively to a
generic fiber $(\GS(b),\ve{p}(b))$ and  the central fiber
$(\GS(b_0),\ve{p}(b_0))$ of this family. Let $e$ be the edge corresponding
to the nodal point that is deformed. We say that $\kBGt$ is
obtained by {\it contracting} the edge $e$ of  $\kBGg$, and to
indicate this we use the notation  $\kBGg < \kBGt$.

\paragraph{Contraction morphism of u-planar trees}
The definition of contraction morphisms of u-planar trees is
the same as that of contraction morphisms of o-planar trees. By contrast,
the contraction of an edge of an u-planar tree is not a well-defined
operation: We can think of a deformation of a real node as
the family $\{x \cdot y = t \mid t \in \R \}$. According to the sign
of the deformation parameter $t$, we obtain two different
unoriented combinatorial types of $\Gs$-invariant curves,
see Figure 1.
 Different u-planar trees
$\BGg_{(i)}$ that are obtained  by contraction of the same edge of
 $\BGt$ correspond to different signs of deformation parameters.

Further details of contraction morphisms for o/u-planar trees can be
found in \cite{c}.
\begin{figure}[ht] 
\begin{center}
\includegraphics[width=75ex]{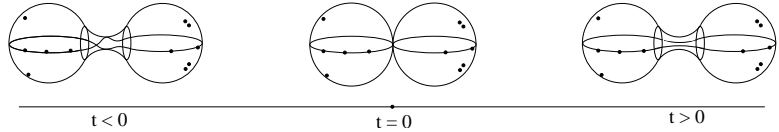}
\caption{Two possible deformations of a real nodal point}. 
\end{center}
\end{figure}

\subsection{Moduli space of $\Gs$-invariant curves}

The moduli space $\cmod{\BS}$ comes equipped with a natural real
structure. The involution
\begin{equation}
\label{eqn_p_r_str}
c: \curve  \mapsto(\overline{\GS};\ve{p})
\end{equation}
 gives the principal real structure of $\cmod{\BS}$.

On the other hand, the permutation group $\S_n$ acts on $\cmod{\BS}$
via relabeling: For each $\Gr \in \S_n$, there is an holomorphic map
$\psi_{\Gr}$ defined by $\curve \mapsto (\GS;\Gr(\ve{p})) := (\GS;
p_{\Gr(s_1)}, \cdots,p_{\Gr(s_n)})$. For each involution $\Gs \in
\S_n$, we have an additional real structure
\begin{equation}
\label{eqn_r_str}
c_{\Gs} := c \circ \psi_{\Gs}: \curve \mapsto
(\overline{\GS};\Gs(\ve{p}))
\end{equation}
of $\cmod{\BS}$. The real part $\rmod{\BS}$ of the real structure
$c_{\Gs}: \cmod{\BS} \to \cmod{\BS}$ gives the moduli space of
$\Gs$-invariant curves:

\begin{thmquote}[Ceyhan, \cite{c}]
\label{thm_r_moduli}

(a) For any $|\BS| \geq 3$, $\rmod{\BS}$ is a smooth projective real
manifold of dimension $|\BS|-3$.

(b) Any  $\Gs$-equivariant family $\pi_B: \comp{U_B} \to \comp{B}$
of  $\BS$-pointed stable curves is induced by a unique pair of real
morphisms
\begin{eqnarray}
\begin{CD}
\comp{U_B}  @> \hat{\kappa} >>   \umod{\BS} \\
@V{\pi_S}VV      @VV{\pi}V   \\
\comp{B}       @> \kappa >>       \cmod{\BS}.          \nonumber
\end{CD}
\end{eqnarray}

(c) Let $\fM_\Gs(\C)$ be the contravariant functor that sends each real
variety $(\comp{B},c_B)$ to the set of $\Gs$-equivariant families of
curves over $\comp{B}$. The moduli functor $\fM_\Gs(\C)$ is represented by the
real variety $(\cmod{\BS},c_\Gs)$.

(d) Let $\real{\fM_\Gs}$ be the contravariant functor that sends each
real analytic manifold $R$ to the set of families of $\Gs$-invariant
curves over $R$. The moduli functor $\real{\fM_\Gs}$ is represented by
the real part $\rmod{\BS}$ of $(\cmod{\BS},c_\Gs)$.
\end{thmquote}

\begin{remark}
The group of holomorphic automorphisms of $\cmod{\BS}$ that respect
its stratification is isomorphic to $\S_n$. Therefore, the real structures
preserving the stratification of $\cmod{\BS}$  are of  the form (\ref{eqn_r_str}) (see \cite{c}).

However, we don't know whether there exist real structures other than
(\ref{eqn_r_str}), since the whole group of holomorphic automorphisms
$Aut(\cmod{\BS})$ is not necessarily isomorphic to $\S_n$. For example,
the automorphism group of $\cmod{\BS}$ is $PSL_2(\C)$ when $|\BS|=4$.

It is believed that $Aut(\cmod{\BS}) \cong \S_n$ for $|\BS| \geq 5$. In fact,
it is true for $|\BS|=5$ and a proof can be found in \cite{dol}.
To the best of our knowledge, there is no systematic exposition of
$Aut(\cmod{\BS})$ for $|\BS| >5$.
\end{remark}

\subsection{A stratification of moduli space $\rmod{\BS}$}
\label{sec_newst}
A stratification for $\rmod{\BS}$ can be obtained by using the stratification 
of $\cmod{\BS}$ given in Theorem \ref{thm_complex_moduli}.

\begin{lemma}
 \label{lem_inv_tree}
Let $\Gg$ and $\overline{\Gg}$ be the dual trees of $\curve$ and
$(\konj{\GS};\Gs(\ve{p}))$ respectively.

(a) If $\Gg$ and $\overline{\Gg}$ are not isomorphic, then the
restriction of $c_{\Gs}$ on the union of complex strata $\cdiv{\Gg}
\bigcup \cdiv{\overline{\Gg}}$ gives a real structure with empty real
part.

(b) If  $\Gg$ and $\overline{\Gg}$ are isomorphic, then the
restriction of  $c_{\Gs}$ on $\cdiv{\Gg}$ gives a real structure whose
corresponding real part $\rdiv{\Gg}$ is the intersection of
$\rmod{\BS}$ with $\cdiv{\Gg}$.
\end{lemma}

%
%

A tree $\Gg$ is called {\it $\Gs$-invariant} if it is isomorphic to
$\bar{\Gg}$.  We denote  the set of $\Gs$-invariant $\BS$-trees  by
$\tree(\Gs)$.

\begin{thmquote}[Ceyhan, \cite{c}] 
\label{cor_strata}
$\rmod{\BS}$ is stratified by real analytic subsets
\begin{equation*}
\rdiv{\Gg} =  
 \prod_{v_r  \in \BV_\Gg^\R} \real{M^{\hat{\Gs}}_{\BF_\Gg(v_r)}} 
\times
\prod_{\{v,\konj{v}\} \subset \BV_\Gg \smin \BV_\Gg^\R} 
\cmodo{\BF_\Gg(v)}
\end{equation*}
where $\Gg \in \tree(\Gs)$ and $\hat{\Gs}$ is the involution
determined by the action of real structure $c_\GS$ on the special
points labelled by $\BF_\Gg(v_r)$ for $v_r  \in \BV_\Gg^\R$.
\end{thmquote}

Although the notion of $\Gs$-invariant trees leads us to a
combinatorial stratification of $\rmod{\BS}$ as given in Theorem
\ref{cor_strata}, it does not give a stratification in terms of
connected strata. For a $\Gs$-invariant $\Gg$, the real part of the
stratum $\rdiv{\Gg}$ has many connected components. A refinement
of this stratification by using the spaces of $\Z_2$-equivariant point 
configurations in the projective line $\projc$.

\subsubsection{$\Z_2$-equivariant  configurations in $\projc$}

Let $z := [z:1]$ be  an affine coordinate on $\projc$. Consider the
upper half-plane $\H^+ = \{z \in \projc \mid \Im{(z)} >0 \}$ (resp.
lower half plane $\H^- = \{z \in \projc \mid \Im{(z)} <0 \}$) as a
half of the $\projc$ with respect to $z \mapsto \bar{z}$, and the
real part $\projr$ as its boundary. Denote by $\H$ the compactified
disc  $\H^{+} \cup \projr$.

The configuration space of $k =|\bconj(\Gs)|/2$ distinct pairs
of conjugate points in $\H^+ \bigsqcup \H^-$ and $l=|\bfix(\Gs)|$
distinct points in $\projr$ is
\begin{eqnarray*}
\cspp{\BS,\Gs} &:=& \{ (z_{s_{1}},\cdots,z_{s_{2k}};x_{r_1},\cdots,x_{r_l}) \mid
                        z_s \in \H^+ \bigsqcup \H^- \ \mathrm{for}\ s \in \BF \smin \BF^\R,\\
               &  &     z_{s} =      z_{s'} \Leftrightarrow s = s',
                        z_{s} = \bar{z}_{s'} \Leftrightarrow s = \Gs(s') \  \& \\
               &  &     x_r \in \projr \ \mathrm{for}\ r \in \bfix(\Gs), \ x_r =  x_{r'}
                        \Leftrightarrow r=r' \}.
\end{eqnarray*}
The number of connected components of $\cspp{\BS,\Gs}$ is
$2^{k} (l-1)!$  when $l \geq 2$, and $2^{k} $ when $l =1$.
 They are all diffeomorphic to each other; natural diffeomorphisms are given by
$\Gs$-invariant relabelings.

The action of  $SL_2(\R)$ on $\H$ is given by
\begin{eqnarray}
SL_2(\R) \times \H \to \H,\ \ (\GL,z) \mapsto \GL(z)
= \frac{az+b}{cz+d},\ \
\GL =  \left(
\begin{array}{cc}
a   &   b \\
c   &   d \\
\end{array}
\right) \in SL_2(\R). \nonumber
\end{eqnarray}
 It induces an isomorphism $SL_2(\R)/ \pm I
\to Aut(\H)$. The  automorphism group $Aut(\H)$ acts on
$\cspp{\BS,\Gs}$ by
\begin{equation*}
\GL: (z_{s_1},\cdots,z_{s_{2k}};x_{r_1},\cdots,x_{r_l}) \mapsto
     (\GL(z_{s_1}),\cdots,\GL(z_{s_{2k}}); \GL(x_{r_1}),\cdots,\GL(x_{r_l})).
\end{equation*}
This action preserves each of the connected components of $\cspp{\BS,\Gs}$.
It is free when $2k+l \geq 3$, and it commutes with
diffeomorphisms given by $\Gs$-invariant relabelings. Therefore,
the quotient space $\cspq{\BS,\Gs} :=\cspp{\BS,\Gs}  / Aut(\H)$ is a
manifold of dimension $2k+l-3$ whose connected components are
pairwise diffeomorphic.

In addition to the automorphisms considered above, there is a
diffeomorphism of $\cspp{\BS,\Gs}$ which is given in affine
coordinates as follows.
\begin{equation*}
e: (z_{s_1},\cdots,z_{s_{2k}}; x_{r_1},\cdots,x_{r_l})  \mapsto
   (-z_{s_1},\cdots,-z_{s_{2k}};-x_{r_1},\cdots,-x_{r_l}).
\end{equation*}
Consider the quotient space $\csp{\BS,\Gs} = \cspp{\BS,\Gs}/(e)$.
The diffeomorphism $e$ commutes  with each $\Gr$-invariant
relabeling and normalizing action of $Aut(\H)$. Therefore, the
quotient space $\csq{\BS,\Gs} := \csp{\BS,\Gs} /Aut(\H)$ is a
manifold of dimension $2k+l-3$, its connected components are
diffeomorphic to the components of $\cspq{\BS,\Gs}$, and, moreover,
the quotient map $\cspq{\BS,\Gs} \to \csq{\BS,\Gs}$ is a trivial
double covering.

\subsubsection{Connected Components of $\rmodo{S}$}

Each connected component of
$\csq{\BS,\Gs}$  is associated to an
unoriented combinatorial type of $\Gs$-invariant curves, and
each unoriented combinatorial type is  given by a one-vertex u-planar
tree $\BGg$. We denote the connected components of $\csq{\BS,\Gs}$
 by  $C_{\BGg}$.

Every $\Z_2$-equivariant point configuration defines a
$\Gs$-invariant curve. Hence, we define
\begin{equation} \label{eqn_diffeo}
\Xi: \bigsqcup_{\BGg : |\BV_{\Gg}|=1} C_{\BGg} \to \rmodo{\BS}
\end{equation}
which maps  $\Z_2$-equivariant point configurations to the
corresponding isomorphisms classes of irreducible $\Gs$-invariant
curves.

\begin{lemma}
\label{lem_conf_sp}

(a) The map $\Xi$ is a diffeomorphism.

(b) Let $|\bconj(\Gs)|=2k $ and $\bfix(\Gs)=l$. The configuration
space $C_{\BGg}$ is diffeomorphic to
\begin{itemize}
\item $((\H^+)^k \smin \GD) \times \R^{l-3}$ when $l >2$,

\item $((\H^+ \smin \{\im\})^{k-1}  \smin \GD )\times \R^{l-1}$
when $l=1,2$,
\end{itemize}
where $\GD$ is the  union of all diagonals where $z_s =z_{s'}$.
\end{lemma}


\subsubsection{Refinement of the stratification}

We associate a product of configuration spaces  $C_{\kBGg_v}$ and
moduli spaces of pointed complex curves $\cmod{\BF_\Gg(v)}$ to each
o-planar tree $\kBGg$:
\begin{eqnarray*}
C_{\kBGg} =
          \prod_{v \in \BV_{\Gg}^\R} C_{\BGg_v} \times
          \prod_{v \in \BV_\Gg^+} \cmodo{\BF_\Gg(v)}.
 \end{eqnarray*}

For each u-planar $\BGg$, we first choose an o-planar representative
$\kBGg$, and then we set $C_{\BGg} := C_{\kBGg}$. Note that $C_{\BGg}$
does not depend on the o-planar representative.

\begin{thmquote}[Ceyhan, \cite{c}]
\label{thm_strata}

(a) $\rmod{\BS}$ is stratified by $C_{\BGg}$.

(b) A stratum $C_{\BGg}$ is contained in the boundary of
$\csqc{\BGt}$ if and only if  $\BGt$ is obtained by contracting an
invariant set of edges of $\BGg$. The codimension of $C_{\BGg}$ in
$\csqc{\BGt}$ is $|\BE_{\Gg}|-|\BE_{\Gt}|$.
\end{thmquote}

\subsection{Graph homology of $\rmod{\BS}$}

In this section, we summarize the results from  \cite{c1}. We a give a 
combinatorial complex generated by the strata of $\rmod{\BS}$ whose 
homology is isomorphic to the homology of $\rmod{\BS}$.

\subsubsection{A graph complex of $\rmod{\BS}$}

Let $\Gs \in \S_n$ be an involution such that  $\bfix (\Gs) \ne \emp$.
We define a graded group
\begin{eqnarray}
\cG_d &=& \left( \bigoplus_{\BGg : |\BE_{\Gg}|=|\BS|-d-3}
 H_{\dim (C_{\BGg})} (\csqc{\BGg}, Q_{\BGg};\Z) \right) / I_d, \\
&=& \left( \bigoplus_{\BGg : |\BE_{\Gg}|=|\BS|-d-3} \Z\ \cls{\BGg}
\right) / I_d
\end{eqnarray}
where $\cls{\BGg}$ are  the (relative) fundamental classes of
the strata $\csqc{\BGg}$ of $\rmod{\BS}$. Here, $Q_{\BGg}$
denotes the union of the substrata of $\csqc{\BGg}$ of codimension
one and higher.

For $|\bconj(\Gs)| <4$, the subgroup $I_d$ (for degree d) is the trivial
subgroup. In all other cases (i.e., for $|\bconj(\Gs)| \geq 4$),  the subgroup
$I_d$ is generated by the following elements.

\subsubsection{The generators of the ideal of the graph complex.}

The following paragraphs $\fR$-1 and $\fR$-2 describe
 the generators of the   ideal of the    graph complex.

\paragraph{$\fR$-1. Degeneration of a real vertex.}

Consider an o-planar representative $\kBGg$ of a u-planar
tree $\BGg$ such that $|\BE_\Gg|=|\BS|-d-5$, and
consider one of its vertices $v \in \BV_\Gg^\R$ with $|v| \geq 5$
and $|\BF^+_\Gg(v)| \geq 2$. Let $f_i,\bar{f}_i
\in \BF_\Gg \smin \BF_\Gg^\R$ be conjugate pairs of flags for
$i=1,2$ such that $f_1,f_2 \in \BF_\Gg^+(v)$ of $\kBGg$, and
let $f_3 \in \BF_\Gg^\R$. Put $\BF = \BF_\Gg(v) \smin
\{f_1,\bar{f}_1,f_2,\bar{f}_2,f_3\}$

We define two u-planar trees $\BGg_1$ and $\BGg_2$ as follows.

The o-planar representative $\kBGg_1$ of $\BGg_1$
is obtained by inserting a pair of conjugate edges $e=(f_{e},f^{e})$,
$\bar{e} = (f_{\bar{e}},f^{\bar{e}})$ into $\kBGg$ at $v$ in
such a way that $\kBGg_1$ gives $\kBGg$ when we contract the edges $e,\bar{e}$.
Let  $\Bdd_{\Gg_1}(e) =\{\tilde{v},v^{e}\}$, $\Bdd_{\Gg_1}(\bar{e}) =
\{\tilde{v},v^{\bar{e}}\}$. Then, the distribution of flags of $\kBGg_1$
 is given by $\BF_{\Gg_1}(\tilde{v}) = \BF_1
\cup \{f_3, f_e,f_{\bar{e}}\}$, $\BF_{\Gg_1}(v^{e}) = \BF_2
\cup \{f_1,f_2,f^e\}$ and $\BF_{\Gg_1}(v^{\bar{e}}) =
\konj{\BF}_2 \cup \{\bar{f}_1,\bar{f}_2,f^{\bar{e}}\}$
where $(\BF_1, \BF_2, \konj{\BF}_2)$ is an equivariant partition of $\BF$.

The o-planar representative $\kBGg_2$ of  $\BGg_2$ is
obtained by inserting a pair of real edges $e_1=(f_{e_1},f^{e_1})$,
$e_2 =(f_{e_2},f^{e_2})$ into $\kBGg$ at $v$ in
such a way that $\kBGg_2$ produces $\kBGg$ when we contract the edges $e_1, e_2$.
Let $\Bdd_{\Gg_2}(e_1) =\{\tilde{v},v^{e_1}\}$,
$\Bdd_{\Gg_2}(e_2)=\{\tilde{v},v^{e_2}\}$.
The sets of flags of  $\kBGg_2$ are $\BF_{\Gg_2}(\tilde{v}) = \ff{\BF}_1 \cup \{f_3 ,
f_{e_1},f_{e_2}\}$, $\BF_{\Gg_2}(v^{e_1}) = \ff{\BF}_2 \cup
\{f_1,\bar{f}_{1},f^{e_1}\}$, $\BF_{\Gg_2}(v^{e_2}) = \ff{\BF}_3
\cup \{f_2,\bar{f}_{2},f^{e_2}\}$
where $(\ff{\BF}_1, \ff{\BF}_2, \ff{\BF}_3)$ is an equivariant partition of $\BF$.


The u-planar trees $\BGg_1, \BGg_2$ are the equivalence classes
represented by $\kBGg_1, \kBGg_2$ given above.

Then, we define
\begin{eqnarray}
\label{eqn_g_relation1}
\cR(\BGg;v,f_1,f_2,f_3) := \sum_{\BGg_1}  \cls{\BGg_1} -
\sum_{\BGg_2}  \cls{\BGg_2},
\end{eqnarray}
where the summation is taken over all possible $\BGg_i,i=1,2$ for a
fixed set of flags $\{f_1,\bar{f}_1,f_2,\bar{f}_2,f_3\}$.

\paragraph{$\fR$-2. Degeneration of a conjugate pair of vertices.}

Consider an o-planar representative $\kBGg$ of a u-planar
tree $\BGg$  such that $|\BE_\Gg|=|\BS|-d-5$, and
a pair of its conjugate vertices $v,\bar{v} \in
\BV_\Gg \smin \BV_\Gg^\R$ with $|v|=|\bar{v}| \geq 4$.
Let $f_i \in \BF_\Gg(v), i=1,\cdots,4$ and   $\bar{f}_i \in
\BF_\Gg(\bar{v})$  be the flags conjugate  to $f_i, i=1,\cdots,4$.
Put $\BF = \BF_\Gg(v) \smin
\{f_1,\cdots,f_4\}$. Let $(\BF_1,\BF_2)$ be a two-partition of
$\BF$,  and $\konj{\BF}_1,\konj{\BF}_2$ be the sets of flags that are conjugate
to the flags in $\BF_1,\BF_2$ respectively.

We define two u-planar trees $\BGg_1$ and $\BGg_2$ as follows.

The o-planar representative $\kBGg_1$ of  $\BGg_1$
is obtained by inserting a pair of conjugate
edges $e=(f_{e},f^{e})$, $\bar{e} =(f_{\bar{e}},f^{\bar{e}})$ to
$\kBGg$ at $v,\bar{v}$ in such a way that $\kBGg_1$ produces
$\kBGg$ when we contract the edges $e, \bar{e}$.
Let $\Bdd_{\Gg_1}(e) = \{v_{e},v^{e}\}$,
$\Bdd_{\Gg_1} (\bar{e}) =\{v_{\bar{e}},v^{\bar{e}}\}$.
 The sets of flags of $\kBGg_1$  are $\BF_{\Gg_1}(v_{e}) = \BF_1 \cup \{f_1,f_2,f_e\}$,
$\BF_{\Gg_1}(v^{e}) = \BF_2 \cup \{f_3,f_4,f^e\}$ and
$\BF_{\Gg_1}(v_{\bar{e}}) = \overline{\BF}_1 \cup
\{\bar{f}_1,\bar{f}_2,f_{\bar{e}}\}$, $\BF_{\Gg_1}(v^{\bar{e}}) =
\overline{\BF}_2 \cup \{\bar{f}_3,\bar{f}_4,f^{\bar{e}}\}$.

The o-planar representative $\kBGg_2$ of   $\BGg_2$ is
also obtained by inserting a pair of conjugate edges into $\kBGg$ at the
same vertices $v,\bar{v}$, but the flags are distributed differently on vertices.
Let  $\Bdd_{\Gg_2}(e) =\{v_e,v^e\}$, $\Bdd_{\Gg_2}(\bar{e}) = \{v_{\bar{e}},v^{\bar{e}}\} $.
Then, the distribution of the flags of $\kBGg_2$
 is given by
$\BF_{\Gg_2}(v_{e}) = \BF_1 \cup \{f_1,f_3,f_e\}$,
$\BF_{\Gg_2}(v^{e}) = \BF_2 \cup \{f_2,f_4,f^e\}$,
$\BF_{\Gg_2}(v_{\bar{e}}) = \overline{\BF}_1 \cup \{\bar{f}_1,
\bar{f}_3,f_{\bar{e}}\}$, and
$\BF_{\Gg_2}(v^{\bar{e}}) =
\overline{\BF}_2 \cup  \{ \bar{f}_2,\bar{f}_4,f^{\bar{e}}\}$.

The u-planar trees $\BGg_1, \BGg_2$ are the equivalence classes
represented by $\kBGg_1, \kBGg_2$ given above.

We define
\begin{eqnarray} \label{eqn_g_relation2}
\cR(\BGg;v,f_1,f_2,f_3,f_4) := \sum_{\BGg_1} \cls{\BGg_1}
-  \sum_{\BGg_2}  \cls{\BGg_2},
\end{eqnarray}
where the summation is taken over all $\BGg_1, \BGg_2$ for a fixed
set of flags $\{f_1,\cdots,f_4\}$.

The ideal $I_d$ is generated by $\cR(\BGg;v,f_1,f_2,f_3)$ and
$\cR(\BGg;v,f_1,f_2,f_3,f_4)$ for all $\BGg$ and $v$ satisfying
the required conditions above.

\subsubsection{The boundary homomorphism of the graph complex}

We define the {\it graph complex} $\cG_\bullet$ of the moduli space
$\rmod{\BS}$ by introducing a boundary map $\dd: \cG_d \to
\cG_{d-1}$
\begin{eqnarray}
\label{eqn_differential} \dd\  \cls{\BGt} \ = \ \sum_{\BGg} \ \pm \
\cls{\BGg},
\end{eqnarray}
where the summation is taken over all u-planar trees $\BGg$ which
give $\BGt$ after contracting one of their real edges.

\begin{thmquote}[Ceyhan, \cite{c1}]
 \label{thm_homology1}
The homology of the graph complex $\cG_\bullet$ is isomorphic to the
singular homology $H_*(\rmod{\BS};\Z)$ for $\bfix(\Gs) \ne \emp$.
\end{thmquote}

\begin{remark}
In  \cite{c1}, the graph homology is defined and a similar theorem
is proved for  $\bfix(\Gs) = \emp$ case. The generators of the ideal
are slightly different in this case.
\end{remark}

\begin{remark}
If $|\BS| >4$ and  $|\bfix(\Gs)| \ne 0$, then the moduli space
$\rmod{\BS}$ is not orientable. A combinatorial construction
of the orientation double covering of $\rmod{\BS}$ is given in
\cite{c}. A stratification of the orientation cover is given in
terms of certain equivalence classes of o-planar trees. By following
the same ideas above, it is possible to construct a graph
complex generated by fundamental classes of the strata that
calculates the homology of the orientation double cover of $\rmod{\BS}$.
\end{remark}

\begin{remark}
In their recent preprint \cite{ehkr}, Etingof {\it et al} calculated the cohomology 
algebra $H^*(\rmod{\BS};\Q)$ in terms of generators and relations for the 
$\Gs = id$ case.  Until this work, little was known about the topology of 
$\rmod{\BS}$ (except  \cite{DJS,de,gm,ka}).

The graph homology, in a sense, treats the homology of the moduli space
$\rmod{\BS}$ in the complementary directions to \cite{ehkr}: The graph complex 
provides a recipe to calculate the homology of $\rmod{\BS}$ in  $\Z$ coefficent 
for all possible  involutions $\Gs$, and it reduces to cellular complex of the moduli 
space $\rmod{\BS}$ for $\Gs=id$ (see \cite{de,ka}). 
Moreover, our presentation is based on stratification of $\rmod{\BS}$
which suits for investigating Gromov-Witten-Welschinger classes. 
\end{remark}

The graph homology provides a set of homological relations between the 
classes of $\rdivc{\BGg}$ that are essential for quantum
cohomology of real varieties:

\begin{corollary}
If $\Gt$ is a $\Gs$-invariant tree, then
\begin{eqnarray} 
\label{eqn_a8}
\dd \ \callr{\Gt} =
\sum_{{\text{all\ possible}} \atop {\text{u-planar\ str.\ of} \ \BGg}} 
\sum_{\BGg} \pm \cls{\BGg}
 = \sum_{\Gg} \callr{\Gg}
\end{eqnarray}
is homologous to zero.
\end{corollary}

\begin{corollary}
If $\BGg$ be an o-planar tree satisfying the condition required in $\fR$-1,
then the sum
\begin{eqnarray}
\label{eqn_cardy}
\sum_{{\text{all\ possible}} \atop {\text{u-planar\ str.\ of} \ \BGg}} \cR(\BGg;v,f_1,f_2,f_3) = 
\sum_{\Gg_1} \callr{\Gg_1} - \sum_{\Gg_2} \callr{\Gg_2} 
\end{eqnarray}
is homologous to zero.
\end{corollary}


\section*{Part II: Quantum cohomology of real varieties}

In this part, we recall the definition of Gromov-Witten classes and 
quantum cohomology. Then, we introduce Welschinger classes 
and quantum cohomology of real varieties. Our  main
goal is to revise Horava's attempt of quantum cohomology of real varieties
by investigating the properties of Gromov-Witten and Welschinger invariants.

\section{Gromov-Witten classes}

Let $X$ be a projective algebraic manifold, and let $\Gb \in H_2(X;\Z)$ such that
$(L\cdot \Gb) \geq 0$ for all K\"ahler $L$. Let $\cmap{\BS}$ be the set of 
isomorphism classes of $\BS$-pointed maps $\map$ where $\GS$ is a projective
nodal curve of genus zero,    $p_{s_1},\cdots,p_{s_n}$ are distinct
smooth labeled  points of $\GS$, and $f:\GS \to X$ is a morphism satisfying 
$f_*([\GS])  = \Gb$.

For each labelled point $s \in \BS$, the moduli space stable maps $\cmap{\BS}$
inherits a canonical {\it evaluation} map
\begin{eqnarray*}
ev_s: \cmap{\BS} \to X 
\end{eqnarray*}
defined for $\map$ in $\cmap{\BS}$ by:
\begin{eqnarray*}
ev_s: (\GS;\ve{p},f) \mapsto f(p_s).
\end{eqnarray*}

Given classes $\mu_{s_1},\cdots,\mu_{s_n} \in H^*(X)$, a product is determined in the
ring $H^*\cmap{\BS}$ by:
\begin{eqnarray}
\label{eqn_prod}
ev^*_{s_1} (\mu_{s_1}) \cup \cdots \cup ev^*_{s_n} (\mu_{s_n}). 
\end{eqnarray}
If $\sum \text{codim}(\mu_s) = \text{dim} (\cmap{\BS})$, then 
the product (\ref{eqn_prod}) can evaluated on the fundamental 
class of $\cmap{\BS}$. In such a case, the {\it Gromov-Witten} invariant
is defined as the degree of the evaluation maps:
\begin{eqnarray*}
\int_{\cmap{\BS}} ev^*_{s_1} (\mu_{s_1}) \cup \cdots \cup ev^*_{s_n} (\mu_{s_n}). 
\end{eqnarray*}
This gives an appropriate counting of parametrized  curves genus zero, 
lying in a given homology class $\Gb$ and satisfying 
certain incidence conditions encoded by $\mu_s$'s.

The above definition of Gromov-Witten invariants leads to a more 
general Gromov-Witten invariants in $H^* \overline{M}_{\BS}$: 
Given classes $\mu_{s_1},\cdots,\mu_{s_n}$,  we have
\begin{eqnarray*}
I_{\BS,\Gb}^X (\mu_{s_1} \otimes \cdots \otimes \mu_{s_n}) = 
\nu_*(ev^*_{s_1} (\mu_{s_1}) \cup \cdots \cup ev^*_{s_n} (\mu_{s_n}))
\end{eqnarray*}
where $\nu: \cmap{\BS} \to \overline{M}_{\BS}$ is the projection that forgets
the morphisms $f: \GS \to X$ and stabilizes resultant pointed curve (if necessary).
The set of multilinear maps 
\begin{eqnarray*}
\{ I^X_{\BS,\Gb}:  \bigotimes_\BS H^*X \to H^* \overline{M}_{\BS} \}
\end{eqnarray*}
is called the  {\it (tree-level) system of Gromov-Witten classes}.

\subsection{Cohomological field theory}

A two-dimensional cohomological field theory (CohFT) with 
coefficient field $\kk$ consists of the following data: 

\begin{itemize}
\item A  $\kk$-linear superspace (of fields) $\cH$, endowed with an even 
non-degenerate  pairing $g$. 

\item A family of even linear maps (correlators) 
\begin{eqnarray}
\label{eqn_cohft}
\{ I_{\BS}:  \bigotimes_\BS \cH \to H^* \overline{M}_{\BS} \}
\end{eqnarray}
defined for all $|\BS| \geq 3$.
\end{itemize}

\noindent
These data must satisfy the following axioms:

\smallskip
\noindent
{\bf A.} {$\S_n$-covariance:}
The maps $I_{\BS}$ are compatible with the actions of $\S_n$ on 
$\bigotimes_\BS \cH$ and on $ \overline{M}_{\BS}$.

\smallskip
\noindent
{\bf B.} {Splitting:} Let $\{\mu_a\}$ denote a basis of $\cH$, 
$\GD := \sum g^{ab} \mu_a \otimes \mu_b$ is the 
Casimir element of the pairing.

 Let $\BS_a=\{s_{a_1},\cdots,s_{a_m}\}$,  $\BS_b=\{s_{b_1},\cdots, s_{b_n}\}$ 
and  $\BS= \BS_1 \sqcup \BS_2$. Let $\Gg$ be an $\BS$-tree such that 
$\BV_\Gg =\{v_a,v_b\}$, and the set of flags $\BF_\Gg (v_a)=\BS_a \cup \{s_a\}$, 
$\BF_\Gg (v_b)= \BS_b \cup \{s_b\}$, and let
\begin{eqnarray*}
\Gvp_{\Gg}: \overline{M}_{\BF_\Gg (v_a)} \times  \overline{M}_{\BF_\Gg (v_b)} \to 
\overline{M}_{\BS}
\end{eqnarray*}
be the map  which assigns to the pointed curves
$(\GS_a;p_{s_{a_1}},\cdots,p_{s_{a_m}},p_{s_a})$ and 
$(\GS_b;p_{s_{b_1}},\cdots,p_{s_{b_n}},p_{s_b})$, their union $\GS_a \cup \GS_b$ 
identified at $p_{s_a}$ and $p_{s_b}$. 

The Splitting Axiom reads:
\begin{eqnarray}
\label{eqn_splitc}
&&\Gvp_{\Gg}^* (I_{\BS}(\mu_{s_1} \cdots  \mu_{s_n}))  \nonumber
     \\
&& = \Ge(\Gg) \sum_{a,b} \ g^{ab} \ 
I_{\BF_\Gg (v_a)}(\bigotimes_{\BS_a} \mu_{s_*} \otimes  \mu_{s_a})) 
\otimes
I_{\BF_\Gg (v_a)}(\bigotimes_{\BS_b} \mu_{s_*} \otimes  \mu_{s_b}))
\end{eqnarray}
where $\Ge(\Gg)$ is the sign of permutation $(\BS_1,\BS_2)$ on $\{\mu_*\}$
of odd dimension.

\bigskip
CohFT's are abstract generalizations of tree-level systems of Gromov-Witten 
classes: any system of GW-classes for a manifold $X$ satisfying 
appropriate assumptions gives rise to the cohomological field theory 
where $\cH = H^*(X;\kk)$.

\subsection{Homology operad of $\overline{M}_{\BS}$}

There is   another useful reformulation of CohFT.  By dualizing
(\ref{eqn_cohft}), we obtain a family of maps
\begin{eqnarray}
\label{eqn_operad1}
\{ \BY_{\BS}:  H_* \overline{M}_{\BS \cup \{s\}} \to Hom(\bigotimes_\BS \cH, \cH) \}.
\end{eqnarray}
Therefore, any homology class in $\overline{M}_{\BS \cup \{s\}}$
is interpreted as an $n$-ary operation on $\cH$. The additive relations given in
Section \ref{sec_additive} and the splitting axiom (\ref{eqn_splitc})  become identities 
between these operations  i.e., $\cH$ carries the structure of an algebra over the 
cyclic operad $H_* \overline{M}_{\BS \cup \{s\}}$.

\begin{figure}[ht] 
\begin{center}
\par\medskip
\includegraphics[width=36ex]{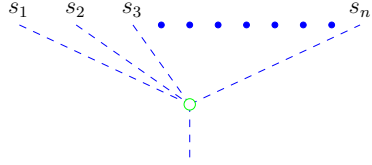} \\
\caption{$n$-ary operation corresponding to the fundamental class of 
$\overline{M}_{\BS \cup \{s\}}$}.
\par\medskip
\end{center}
\end{figure}

\subsubsection{$Comm_{\infty}$-algebras}

Each $n$-ary operation is a composition of $k$-ary operations corresponding to
fundamental classes of $ \overline{M}_{\BF \cup \{s\}}$ for some $\BF$ of order
$k \leq n$. Reprehasing an algebra over $H_* \overline{M}_{\BS \cup \{s\}}$-operad
by using additive relations in $H_* \overline{M}_{\BS \cup \{s\}}$ provides
us a {\it $Comm_{\infty}$-algebra}.

The structure of {\it $Comm_\infty$-algebra} on $(\cH,g)$
is a sequence of even polylinear maps $\BY_k: \cH^{\otimes k} \to A,
k \geq 2$, satisfying the following conditions:

\smallskip
\noindent
{\bf A$'$.} {Higher commutativity:} $\BY_{k}$ are $\S_k$-symmetric.

\smallskip
\noindent
{\bf B$'$.} {Higher associativity:} For all $m \geq 0$, and 
$\mu_{\Ga}, \mu_{\Gb}, \mu_{\Gd}, \mu_{i_1}, \cdots, \mu_{i_m} \in A$, we have
\begin{eqnarray*}
\sum_{{\Gt: \BS_1 \sqcup \BS_2 = \nn} \atop {\Ga,\Gb,j_* \in \BS_1 \& \Gd, k_* \in \BS_2}}  
(-1)^{\Ge(\Gt)} \BY_{|\BS_2|} (\BY_{|\BS_1|} (\mu_\Ga, \mu_\Gb, \mu_{j_1},\cdots, \mu_{j_r}), \mu_\Gd,
\mu_{k_1}, \cdots, \mu_{k_s}) \\
= \sum_{{\Gg: \BS'_1 \sqcup \BS'_2 = \nn} \atop {\Gb,\Gd,j_* \in \BS'_2 \& \Ga, k_* \in \BS'_1}}  
(-1)^{\Ge(\Gg)} \BY_{|\BS'_2|} (\BY_{|\BS'_1|} (\mu_\Gb, \mu_\Gd, \mu_{j_1}, \cdots, \mu_{j_r}),
\mu_\Ga, \mu_{k_1}, \cdots, \mu_{k_s}) 
\end{eqnarray*}
Here, the summations runs over all partitions of $\nn$ into two disjoint subsets
$(\BS_1,\BS_2)$  and $(\BS'_1,\BS'_2)$ satisfying required conditions. 
Note that, these relations are
consequences of the homology relations given in Theorem \ref{thm-ring}.

Since we identify the bases with $n$-corollas of the cyclic operad 
$H_* \overline{M}_{\BS \cup \{s\}}$ (see, Figure 2), it is represented by as a linear span of all possible  $\BS'$-trees
modulo the higher associativity relations for $\BS' = \BS \cup \{s\}$:

\begin{figure}[ht] 
\begin{center}
\par\medskip
\includegraphics[width=75ex]{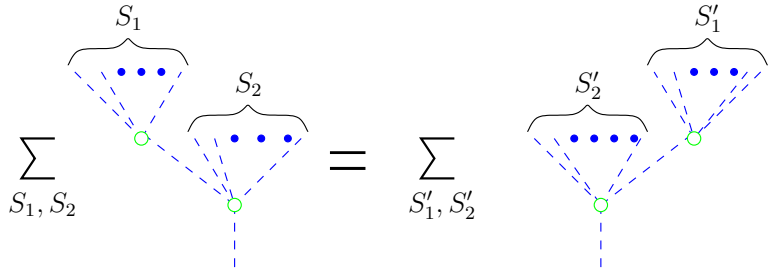} \\
\caption{Higher associativity relations}. 
\par\medskip
\end{center}
\end{figure}

\noindent
We call  $Comm_\infty$-algebra $(\cH,g,\BY_*)$ {\it cyclic} if the tensors
\begin{eqnarray*}
&\BGe_{k+1}: \cH^{\otimes k+1} \to \kk,& \\ 
&\BGe_{k+1} (\mu_1, \cdots, \mu_k, \mu_{k+1})& := 
g (\BY_k(\mu_1,  \cdots, \mu_k), \mu_{k+1})
\end{eqnarray*}
are $\S_{k+1}$-symmetric .

\subsection{Gromov-Witten potential and quantum product}
Let $X$ be a manifold equipped with a system of tree level 
GW-classes.  Put
\begin{eqnarray*}
\langle \mu_{s_1} \cdots \mu_{s_n} \rangle_\Gb
=  \langle I_{\BS,\Gb}^X \rangle (\mu_{s_1} \otimes \cdots  \otimes \mu_{s_n})
:= \int_{ \overline{M}_\BS}  I_{\BS,\Gb}^X (\mu_{s_1} \otimes \cdots  \otimes \mu_{s_n}).
\end{eqnarray*}

\noindent
We define $\GP$ as a formal sum depending 
on a variable point $\GG \in H^*(X)$: 

\begin{eqnarray*}
\GP(\GG) :=  \sum_{|\BS| \geq 3} \sum_{\Gb}
\langle I_{\BS,\Gb}^X \rangle (\GG^{\otimes |\BS|}) \frac{q^\Gb}{|\BS| !}. 
\end{eqnarray*}

The quantum multiplication is defined a product on tangent space
of homology space $H^*(X)$ by 
\begin{eqnarray*}
\dd_a * \dd_b = \sum_{c,d} \GP_{abc} \ \dd_d
\end{eqnarray*}
where $\GP_{abc}$ denotes the third derivative $\dd_a \dd_b \dd_c (\GP)$
of the potential function.

\begin{thmquote}[Kontsevich \& Manin, \cite{km1}]
This definition makes the tangent sheaf of the homology space
$H^*(X,\C)$ a Frobenius manifold. 
\end{thmquote}

The main property, the associativity of
quantum product is a consequence of splitting axiom of GW-classes and the 
defining relations of homology ring of $\overline{M}_\BS$ 
(see, Theorem \ref{thm-ring}).

\section{Gromov-Witten-Welschinger classes}

In this section, we introduce the Welschinger classes by following
the same principles  and steps which lead to Gromov-Witten theory: 
Firstly, we introduce the
real enumerative invariants in terms of homology of moduli space 
of suitable maps. Then, we transfer this definition to a new one in
terms of the homology of $\rmod{\BS}$. We then further generalize
our definition to open-closed CohFT's and study these algebraic
structures by using $H_* \rmod{\BS}$ i.e., introduce the quantum
cohomology of real varieties.

\subsection{Moduli space of stable real maps}
Let $(\comp{X},c_X)$ be a projective real  algebraic manifold.
Let $ \Bd := (\Gb, d) \in H_2(\comp{X}) \oplus H_1(\real{X})$.

We call a stable map $\map \in \cmap{\BS}$  {\it $\Gs$-invariant} if 
 $\GS$ admits a real structure $c_\GS:\GS \to \GS$ such that 
$c_\Gs (p_s) = p_{\Gs(s)}$, $c_X \circ f = f \circ c_\GS$ and 
$f_*([\real{\GS}]) = \pm d \in H_1(\real{X})$.

We denote the moduli space of $\Gs$-invariant stable maps by $\rmap{\BS}$. 
The open stratum $\rmapo{\BS}$ of $\rmap{\BS}$ is a subspace  of $\rmap{\BS}$
where domain curve $\GS$ is irreducible.  It is important to note that the 
moduli space $\rmap{\BS}$ has boundaries.  The $\Gs$-invariant maps $\map$ 
lying in the
boundaries have at least two components of domain curve $\GS$ 
which  are not contracted to a point by  morphism $f: \GS \to X$.

Obviously, $\rmap{\BS}$ is a subspace of the real part of the real structure
\begin{eqnarray*}
c_\Gs: \cmap{\BS} \to \cmap{\BS}; \ \ \  \
\map \mapsto (\overline{\GS}; \Gs(\ve{p});c_X \circ f).
\end{eqnarray*}
Therefore, the restrictions of the evaluation maps provide us 
\begin{eqnarray*}
\begin{array}{ll}
ev_s: \rmap{\BS} \to \comp{X}, & \ \text{for}\  s \in \bconj(\Gs), \\
ev_s: \rmap{\BS} \to \real{X}, & \ \text{for}\  s \in \bfix(\Gs).
\end{array}
\end{eqnarray*}

These moduli spaces have been extensively studied in the open 
Gromov-Witten invariants context. The compactification of these
moduli spaces studied in  (see \cite{fu,liu,s}).  The 
orientibility of the open stratum $\rmapo{\BS}$ has been noted  first 
by Fukaya {\it et al} (for $\Gs =id$ case), Liu and (implicitly) by 
Welschinger in \cite{fu,liu, w1,w2}.
Solomon treated this problem in most general setting (see \cite{s}).

\subsection{Welschinger invariants as degrees of evalution maps}

In a series of papers \cite{w1}-\cite{w4},  Welschinger defined a set of  invariants 
counting, with appropriate weight $\pm 1$, real rational $J$-holomorphic curves  
intersecting a generic $\Gs$-invariant 
collection of marked points. Unlike the usual homological definition of Gromov-Witten 
invariants, Welschinger invariants are originally defined  by assigning signs to 
individual curves based on certain geometric-topological criteria. A homological
interpretation of Welschinger invariants has been given by J. Solomon very 
recently (see \cite{s}):

Let $H^*(X;\det (T\real{X}))$ denote the cohomology of $\real{X}$ with 
coefficients in the flat line bundle $\det (T\real{X})$.
Let $\Ga_s \in \GO^*(\real{X}, T\real{X})$ be representatives of a set 
classes in  $H^*(X;\det (T\real{X}))$ for $s \in \bfix(\Gs)$. Furthermore,
for $\{s,\bar{s}\} \subset \bconj(\Gs)$, $\mu_s \in \GO^* \comp{X}$ represent a set of 
 classes in  $H^*(\comp{X})$.
Then, a product is determined in $\GO^* \rmapo{\BS}$ by
\begin{eqnarray}
\label{eqn_form}
\bigwedge_{\{s,\konj{s}\} \subset \bconj(\Gs)} ev_s^* (\mu_s)
\ \wedge \ 
\bigwedge_{s \in \bfix(\Gs)} ev_s^*(\Ga_s).
\end{eqnarray}
If $\sum \text{codim} (\Ga_*) + \sum \text{codim}(\mu_*)= \dim(\rmapo{\BS})$, the
form (\ref{eqn_form}) extends along the critical locus of evaluation map except 
$\Gs$-invariant maps with reducible domain curve. In other words, 
it provides a relative cohomology class in $H^*(\rmap{\BS},\sD)$. Here, $\sD$ is the subset of 
$\rmap{\BS}$ whose elements $\map$ have more than one components. 
This extension of the differential form (\ref{eqn_form}) is a direct consequence of Welschinger's theorems.

If $\mu_*$ and  $\Ga_*$ are Poincare duals of point classes (respectively in 
$\comp{X}$ and $\real{X}$), then one can define
\begin{eqnarray}
N^{\Gs}_{\BS,\Bd}   := \int_{[\rmap{\BS}]} \left\{
\bigwedge_{\{s,\konj{s}\} \subset \bconj(\Gs)} ev_s^* (\mu_s)
\ \wedge \ 
\bigwedge_{s \in \bfix(\Gs)} ev_s^*(\Ga_s) \right\}.
\end{eqnarray}

\begin{proposition} (Solomon, \cite{s})
The sum
\begin{eqnarray}
\sum_{d} N^{\Gs}_{\BS,\Bd}
\end{eqnarray}
which is taken over all possible homology classes $d \in H_1(\real{X})$  realized
by $\Gs$-invariant maps, is  equal to Welschinger invariants.
\end{proposition}

\subsection{Welschinger classes}

Let $\nu: \rmap{\BS} \to \rmod{\BS}$ be the restriction of the
contraction map $\nu: \cmap{\BS} \to \konj{M}_{\BS}$. 
Let $\fD$ be the subspace of $\rmod{\BS}$ whose elements $\curve$
has at least two irreducible components.  Note that, the contraction morphism
$\nu$ maps $\sD \subset \rmap{\BS}$ onto $\fD$.

By using the contraction morphism as in the definition of 
Gromov-Witten classes, Solomon's definition of Welschinger invariants can be 
put into a more general setting:
\begin{eqnarray*}
W^X_{\BS,\Bd} ( 
\bigotimes_{ \{s,\bar{s}\} \subset \bconj(\Gs)} \mu_{s} 
&\otimes&
\bigotimes_{s \in \bfix(\Gs)} \Ga_{s}) \\ 
& :=&
\nu_* (
\bigwedge_{\{s,\konj{s}\} \subset \bconj(\Gs)} ev_s^* (\mu_s) 
\ \wedge \ 
\bigwedge_{s \in \bfix(\Gs)} ev_s^*(\Ga_s)).
\end{eqnarray*}

The set of multilinear maps
\begin{eqnarray*}
W^X_{\BS,\Bd}: 
\bigotimes_{\bconj(\Gs)/\Gs} H^*(\comp{X}) 
\bigotimes_{\bfix(\BS)} H^*(\real{X}; \det(T\real{X}))
\to 
H^*(\rmod{\BS},\fD)
\end{eqnarray*}
is called the {\it system  Welschinger classes}.

\subsection{Open-closed cohomological fields theory}

An open-closed CohFT with coefficient field $\kk$ consists of the following data: 

\begin{itemize}
\item A pair of superspaces (of closed and open states) $\cH_c$ and $\cH_o$ 
endowed with even non-degenerate pairings $\eta_c$ and $\eta_o$ respectively.


\item Two sets of linear maps (open \& closed  correlators) 
\begin{eqnarray}
\label{eqn_ocohft}
&\{W_\BS^\Gs:& 
\bigotimes_{\bconj(\Gs)/\Gs} \cH_c
\otimes
\bigotimes_{\bfix(\Gs)} \cH_o 
\to H^*(\rmod{\BS},\fD) \} \\
&\{I_\BS: &
\bigotimes_{\BS} \cH_c \to H^*(\konj{M}_\BS) \} \nonumber
\end{eqnarray}
defined for all $|\BS| \geq  3$.
\end{itemize}

\noindent
These data must satisfy the following axioms:

\smallskip
\noindent
{\bf A$'$.} {CohFT of closed states:} The set of maps $\{I_\BS: 
\bigotimes_{\BS} \cH_c \to H^*(\konj{M}_\BS) \}$ forms a CohFT.

\smallskip
\noindent
{\bf B$'$.} {Covariance:} 
The maps $W_{\BS}^\Gs$ are compatible with the actions of $\Gs$-invariant
relabelling on $\bigotimes_{\bconj(\Gs)/\Gs} \cH_c \otimes  \bigotimes_{\bfix(\Gs)} \cH_o$ 
and on $(\rmod{\BS},\fD)$.

\smallskip
\noindent
{\bf C$'$.} {Splitting 1:} 
 Let $\{\Ga_a\}$ denote a basis of $\cH_o$, and let $\GD_o$ be 
 $\sum \eta_o^{ef} \Ga_e \otimes \Ga_f$.

Let $\Gg$ be a $\Gs$-invariant tree with $\BV_\Gg = \BV_\Gg^\R = \{v_e,v_f\}$,
and let $\BF_\Gg(v_e) = \BS_e \cup \{s_e\}$, $\BF_\Gg(v_f) = \BS_f \cup \{s_f\}$
and $\BS = \BS_e \cup \BS_f$. We denote the restriction of $\Gs$ onto 
$\BS_e,\BS_f$ by $\Gs_e,\Gs_f$ respectively. Let
\begin{eqnarray*}
\Gp_\Gg: \rdivc{\Gg} := 
\konj{M}^{\Gs_e}_{\BF_\Gg(v_e)}(\R) \times \konj{M}^{\Gs_f}_{\BF_\Gg(v_f)}(\R)
\hookrightarrow \rmod{\BS}
\end{eqnarray*}
be the embedding of the real divisor $\rdivc{\Gg}$.

The Splitting Axiom reads:
\begin{eqnarray*}
\Gp_\Gg^*(
W_\BS^\Gs
(\bigotimes_{\bconj(\Gs)/\Gs} \mu_*
&\otimes& 
\bigotimes_{\bfix(\Gs)} \Ga_* 
)) \\
= \Ge(\Gg) \sum_{a,b}  &\eta_o^{ef}&   \
W^{\Gs_e}_{\BF_\Gg(v_e)} 
(\bigotimes_{\bconj(\Gs_e)/\Gs_e} \mu_*
\bigotimes_{\bfix(\Gs_e) \smin \{s_e\}} \Ga_* 
\otimes  \Ga_e ) \\
&\otimes & 
W^{\Gs_f}_{\BF_\Gg(v_f)} 
( \bigotimes_{\bconj(\Gs_f)/\Gs_f} \mu_*
\bigotimes_{\bfix(\Gs_f) \smin \{s_f\}} \Ga_* 
\otimes \Ga_f). 
\end{eqnarray*}

\smallskip
\noindent
{\bf D$'$.} {Splitting 2:} 
Let $\Gg$ be a $\Gs$-invariant tree with $\BV_\Gg =\{v_r,v,\bar{v}\}$
and  $\BV_\Gg^\R = \{v_r\}$, and let $\BF_\Gg(v_r) = \BS_r \cup \{s_e,\bar{s}_e\}$, 
$\BF_\Gg(v) = \BS_f \cup \{s_f\}$,  $\BF_\Gg(\bar{v}) = \konj{\BS}_f \cup \{\bar{s}_f\}$
and $\BS = \BS_r \cup \BS_f \cup  \konj{\BS}_f $. We denote the restriction of $\Gs$ onto 
$\BS_r$ by $\Gs_r$. Let
\begin{eqnarray*}
\Gp_\Gg: \rdivc{\Gg} := 
\konj{M}^{\Gs_r}_{\BF_\Gg(v_r)}(\R) \times \konj{M}_{\BF_\Gg(v)}(\C)
\hookrightarrow \rmod{\BS}
\end{eqnarray*}
be the embedding of the subspace $\rdivc{\Gg}$.

Then, the Splitting Axiom reads:
\begin{eqnarray*}
&&\Gp_\Gg^*(
W_\BS^\Gs
(\bigotimes_{\bconj(\Gs)/\Gs} \mu_*
\bigotimes_{\bfix(\Gs)} \Ga_* \otimes)) 
= \Ge(\Gg)  \\
&& \sum_{a,b} \ \eta_c^{ab} \  \
W^{\Gs_r}_{\BF_\Gg(v_r)}
 (\mu_a
\bigotimes_{\bconj(\Gs_r)/\Gs_r} \mu_*
\bigotimes_{\bfix(\Gs_r)} \Ga_* \otimes ) 
\otimes 
I_{\BF_\Gg(v)} (\mu_b \otimes \bigotimes_{\BS_f} \mu_*). 
\end{eqnarray*}

Open-closed CohFTÕs are abstract generalizations of  systems of Gromov- 
Witten-Welschinger classes: any system of GW-classes  and Welschinger classes
for a real variety $(X,c_X)$ satisfying appropriate assumptions gives rise to the 
open-closed CohFT  where  $\cH_c = H^*\comp{X}$ and 
$\cH_o = H^*(\real{X};\det(T\real{X}))$.

\subsection{Quantum cohomology of real varieties: A DG-operad}

In this section, we first give a brief account of Horava's attempt of
defining quantum cohomology of real varieties. Then, we define
quantum cohomology for real varieties as a DG-operad of the
{\it reduced graph complex} of  the moduli space $\rmod{\BS \cup \{s\}}$.
The definition of this DG-operad is based of Gromov-Witten-Welschinger 
classes.

\subsubsection{Quantum cohomology of real varieties: Horava's approach}

The quantum cohomology for real algebraic varieties has been introduced 
surprisingly early, in 1993, by P. Horava in \cite{ho}. In his paper,
Horava describes a $\Z_2$-equivariant topological sigma model on a
real variety $(X,c_X)$ whose set of physical observables in the is  a direct sum 
of the cohomologies of $\comp{X}$ and $\real{X}$:
\begin{eqnarray}
\label{eqn_qcoh}
\cH_c \oplus \cH_o := H^*(\comp{X}) \oplus H^*(\real{X}).
\end{eqnarray}

In the case of usual  topological sigma models, the OPE-algebra is 
known to give a deformation of the cohomology ring of $\comp{X}$. 
One may thus wonder what the classical structure of $\cH_c \oplus \cH_o$ 
is, of which the quantum OPE-algebra can be expected to be a 
deformation. Notice first that there is a natural structure of an 
$\cH_c$-module on $\cH_c \oplus \cH_o$. Indeed, to define a product of a 
cohomology class $\Ga$ of $\real{X}$ with a cohomology class $\mu$ 
of $\comp{X}$, we will pull back the cohomology class $\mu$ by 
$i: \real{X} \hookrightarrow \comp{X}$, and take the wedge 
product with $\Ga$. Equipped with this natural structure, 
the module (\ref{eqn_qcoh}) associated with the pair represented 
by a manifold $\comp{X}$ and a real structure of it, can be chosen as a 
(non-classical) equivariant cohomology ring in the sense of \cite{br}.
Horava required this equivariant cohomology theory to be recovered 
in the classical limit of the OPE-algebra of the equivariant topological 
sigma model.\footnote{Note the important point that the equivariant 
cohomology theory considered in Horava's study is not the same as 
the Borel's  $G$-equivariant cohomology theory based on the classifying spaces
$BG$ of $G$. It is quite evident that classical equivariant cohomology
theory wouldn't provide that rich structure since the classifying space of
$\Z_2$ is $\P^\infty(\R)$ i.e., it mainly contains two torsion elements}

By ignoring the degeneracy phenomena, the correlation functions of such 
a model count the number of equivariant holomorphic curves $f: \GS \to \comp{X}$
meeting a number of subvarieties of  $\comp{X}$ and $\real{X}$ in images of 
prescribed points of $\GS$. The analogue of the quantum cohomology ring 
is now a structure of $H^*(\comp{X})$-module structure on $H^*(\comp{X}) \oplus 
H^*(\real{X})$ deforming cup product.

However, the degeneracy phenomena is the key ingredient which encodes the 
relations that deformed product structure should respect. In the following
paragraph, we define quantum cohomology of real variety as a DG-operad
in order to recover these relations.

\subsubsection{Reduced graph complex}
Let $\cG_\bullet$ be the graph complex of $\rmod{\BS}$. 
The reduced graph complex $\cC_\bullet$ of the moduli space
$\rmod{\BS}$ is a subcomplex of $\cG_\bullet$ given as
follows:
\begin{itemize}
\item The graded group is 
\begin{eqnarray}
\cC_d = \left( \bigoplus_{{\Gg \ \text{are}\ \Gs-\text{invariant}} \atop  
                                            { \& |\BE_{\Gg}|=|\BS|-d-3}} \Z\ \callr{\Gg}
\right) / \cI_d
\end{eqnarray}
where $\callr{\Gg}$ are  the (relative) fundamental class of
the subspaces $\rdivc{\Gg}$ of $\rmod{\BS}$. The ideal of relations
$ \cI_d$ is generated by (\ref{eqn_cardy}).

\item The boundary homomorphism of the reduced graph complex
is given by the restriction of the boundary homomorphism 
$\dd: \cG_d \to \cG_{d-1}$ of the graph complex $\cG_\bullet$ 
(see \ref{eqn_a8}). 
\end{itemize}

\begin{remark}
There is an obvious monomorphism $H_*(\cC_\bullet) \to 
H_*(\cG_\bullet)$.  
A priori, the dualization of (\ref{eqn_ocohft}) requires rather whole 
homology group $H_*(\cG_\bullet)$ than a subgroup $H_*(\cC_\bullet)$.
However, the covariance property of $W_\BS^\Gs$ allows us 
restrict ourselves to subspaces $\rdivc{\Gg}$ instead of the strata
considered in Theorem \ref{thm_strata}. 
\end{remark}

\subsubsection{DG-operad of reduced graph homology}

Let $\cC_\bullet$ be the reduced graph complex of $\rmod{\BS \cup \{s\}}$.
By dualizing (\ref{eqn_ocohft}), we  obtain a family of maps 
\begin{eqnarray}
\label{eqn_swoperad}
\{ \BZ_\BS^\Gs:  \cC_\bullet  \to 
Hom(\bigotimes_{\bconj(\Gs)/\Gs} \cH_c
\bigotimes_{\bfix(\Gs)} \cH_o, \cH_o) \}
\end{eqnarray}
along with an algebra over $H_* \konj{M}_{\BS \cup \{s\}}$-operad 
\begin{eqnarray}
\label{eqn_swoperad+}
\{ \BY_\BS:  H_* \konj{M}_{\BS \cup \{s\}} \to 
Hom(\bigotimes_{\BS} \cH_c, \cH_c) \}.
\end{eqnarray}

Therefore, the homology class of subspace $\rdivc{\Gg}$  in $\rmod{\BS \cup \{s\}}$
is interpreted as an $(k,l)$-ary operation on $\cH_o \oplus \cH_c$
where $k = |\bconj(\Gs)|/2$ and $l = |\bfix(\Gs)|$.

\begin{figure}[ht]
\begin{center}
\par\medskip
\includegraphics[width=56ex]{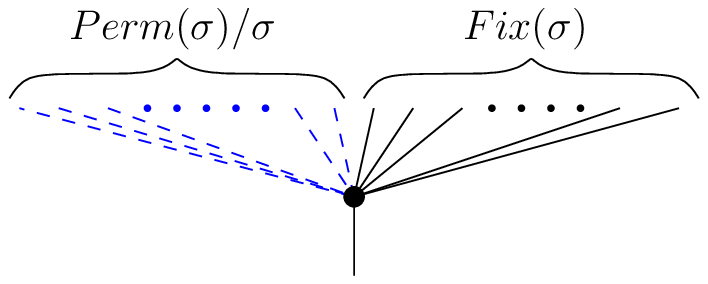} \\
\caption{$(k,l)$-ary operation corresponding to the fundamental class of 
$\rmod{\BS \cup \{s\}}$}.
\par\medskip
\end{center}
\end{figure}

\begin{remark}
The $\Gs$-invariant trees which are used in Lemma \ref{lem_inv_tree}
are different than the trees depict $(k,l)$-ary operation. These new
trees are in fact are quotients of $\Gs$-invariant tree with respect to
their symmetries arising from real structures of the corresponding
$\Gs$-invariant curves.
\end{remark}

In addition to the higher associativity $\BY_\BS$, the additives relations in 
$\cC_\bullet$ require additional conditions on $(k,l)$-ary operations.
The families of maps $\BZ_\BS^\Gs$ and $\BY_\BS$
satisfy the following conditions:

\smallskip
\noindent
{\bf 1.} {\it Higher associativiy relations:} $(\cH_c,\eta_c,\BY_*)$ form a 
$Comm_\infty$-algebra structure.

\smallskip
\noindent
{\bf 2.} {\it $A_\infty$-type of structure:} The additive relations arising from  
the image of the boundary homomorphism $\dd: \cC_* \to \cC_{*-1}$ 
(i.e., from the identity  (\ref{eqn_a8})) and the splitting axiom ({\bf C$'$})  
become the following identities between $(k,l)$-ary operations:

\begin{figure}[ht]
\begin{center}
\par\medskip
\includegraphics[width=70ex]{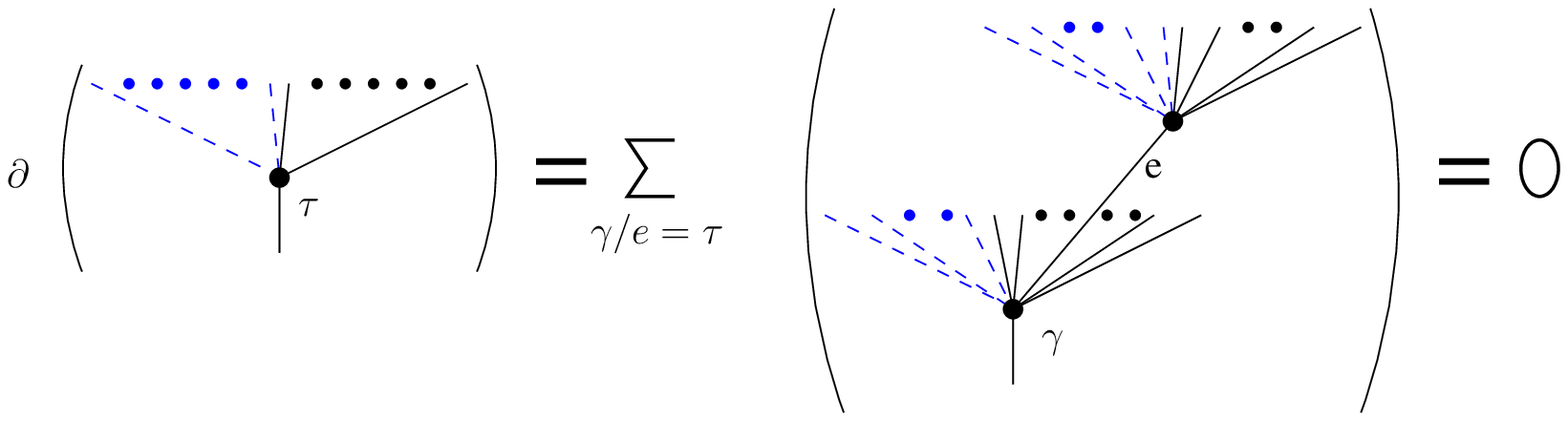} \\
\caption{The image of boundary map provides $A_\infty$-type of relations}.
\par\medskip
\end{center}
 \end{figure}

\noindent
It is easy to see that $(A_o,\eta_o,\BZ_\BS^\Gs)$ reduces to an $A_\infty$-algebra
when $\Gs=id$.

\smallskip
\noindent
{\bf 2.} {\it Cardy type of relations:} The additive relations arising from  
the ideal  $\cI_d$  of the graph complex $\cC_\bullet$ (i.e., from the identity  
(\ref{eqn_cardy})) 
and the splitting axiom ({\bf D$'$})  become the following identities between 
$(k,l)$-ary operations:

\begin{figure}[ht]
\begin{center}
\par\medskip
\includegraphics[width=76ex]{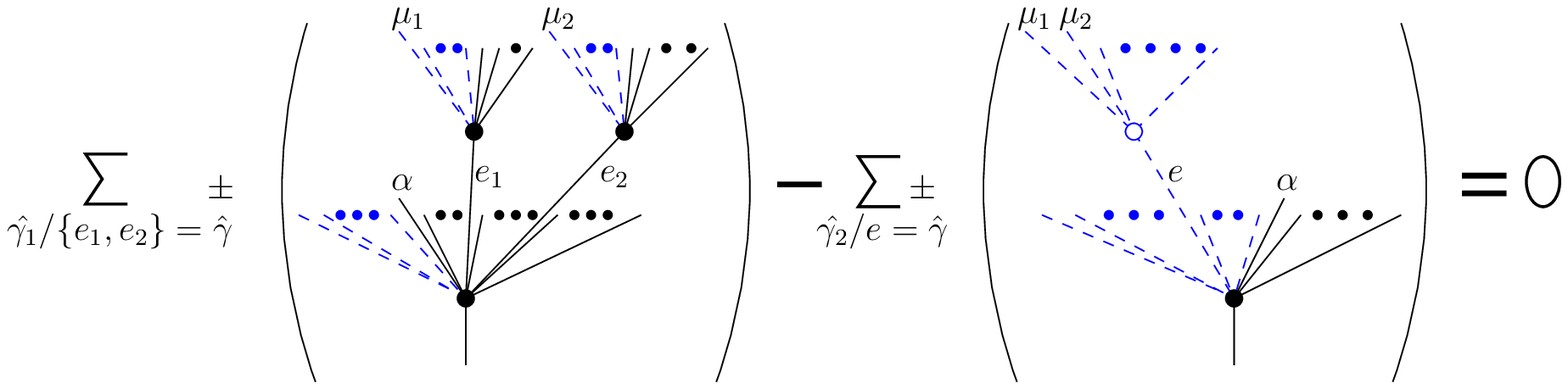} \\
\caption{Cardy type of relations arising from the ideal $\cI_*$}.
\par\medskip
\end{center}
 \end{figure}

Here,  $\hat{\Gg}_1. \hat{\Gg}_2$ are the quotients of $\Gg_1,\Gg_2$ in 
 (\ref{eqn_cardy}) with respect to their $\Z_2$-symmetries induced by real
 structures of corresponding $\Gs$-invariant curves. In this picture, $\mu_1,
 \mu_2$ correspond to the flags $f_1,f_2$, and $\Ga$ corresponds to the real 
 flag $f_3$.

\begin{remark}
The (partial) DG-operad defined above has close relatives which appeared
in the literature.

The first one is  Swiss-cheese operad
which is introduced by Voronov in \cite{vo}.   In fact,  Swiss-cheese
operads are the chain operad of the spaces of $\Z_2$-equivariant 
configuration in $\projc$ (see, Section \ref{sec_newst}).

Kajiura and Stassheff recently introduced {\it open-closed homotopy algebras} 
which are amalgamations of $L_\infty$-algebras with $A_\infty$-algebras, 
see \cite{ks1}-\cite{ks3}. In their setting, the compactification of the configurations
spaces are different than ours, therefore, the differential and the corresponding 
relations are different. In particular, there is no Cardy type of relation in their setting.

Very recently,  Merkulov studied  `operad of formal homogeneous space' in 
\cite{mer}. His approach is motivated by studies on relative obstruction theory, 
its applications (see \cite{r1,r2}). It seems that a version of mirror symmetry
which takes the real structures into account, has connection with this approach. 
\end{remark}

\section*{Part III: Yet again, mirror symmetry!}

In his seminal paper \cite{kon}, Kontsevich proposed that mirror symmetry 
is a (non-canonical) equivalence between the bounded derived category 
of coherent sheaves $D^b Coh(Y)$ on a complex variety $Y$ and the derived 
Fukaya category  $D Fuk(X)$ of its mirror, a symplectic manifold $(X,\Go)$.
This first category consists of chain complexes of holomorphic bundles, with
quasi-isomorphisms and their formal inverses.  Roughly speaking, the derived 
Fukaya category should be constructed from Lagrangian submanifolds 
$L \subset X$ (carrying flat $U(1)$-connection $A$). Morphisms are given by
Floer cohomology of Lagrangian submanifolds. 
One important feature is that, whereas the mirror of a Calabi-Yau 
variety   is another Calabi-Yau variety, the mirrors of Fano varieties  are 
Landau-Ginzburg models; i.e., affine varieties equipped with a map to called the 
superpotential.

Kontsevich also conjectured that the equivalence of derived categories should
imply numerical predictions: For complex side (B-side), consider the diagonal 
subvariety $\GD_Y \subset Y \times Y$, and define the Hochschild cohomology of 
$Y$ to be the endomorphism algebra of $\GD_Y$ regarded as an object of 
$D^b Coh(X \times X)$ \cite{kon}. Kontsevich interpreted the last definition of 
$HH^*(\cO_Y)$ ($\cong H^*(Y, \bigwedge^* TY)$) as computing the space of 
infinitesimal deformations of the bounded derived category of coherent sheaves 
on $Y$ in the class of $A_\infty$-categories. On the other hand, for symplectic
side (A-side), the diagonal $\GD_X \subset X \times X$ is a Lagrangian submanifold
of $(X \times X, (\Go,-\Go))$. The Floer cohomology of the diagonal is canonically
isomorphic to $H^*(X)$. Roughly, above picture suggests that deformation of 
$H^*(X)$ constructed using Gromov-Witten invariants should correspond to 
variations of Hodge structure in B-side.

\subsection*{`Realizing' mirror symmetry}

Let $c_X:X \to X$ be an anti-symplectic involution of $(X,\Go)$.
To recover more about real geometry from Kontsevich's conjecture,
we consider the structure sheaf $\cO_Y$. Calculating its self-Hom's
\begin{eqnarray*}
Ext^i(\cO_Y,\cO_Y) \cong H^{0,i}(Y)
\end{eqnarray*}
shows that if $\cO_Y$ is mirror to the Lagrangian submanifold 
$\real{X} = \fix(c_X)$, 
then we must have
\begin{eqnarray*}
HF^*(\real{X},\real{X}) \cong  H^{0,*}(Y)
\end{eqnarray*}
as graded vector spaces.

Kontsevich conjecture above and the description of quantum cohomology of real
varieties that we have discussed in previous section, suggest us a correspondence 
between two `open-closed homotopy algebras' (in an appropriate sense):

\begin{center}
\begin{tabular}{c c c } 
\ \ \ \ \ Symplectic side $X$ \ (A-side) \ \ \ \ \ 
& $\Longleftrightarrow $   & 
\ \ \ \ \ Complex side $Y$ \ (B-side) \ \ \ \ \   \\
\hline \hline
\rule{0pt}{4ex}
$(\cH_c, \cH_o) =$
& $\Leftrightarrow $ & 
$(\widehat{\cH}_c, \widehat{\cH}_o) = $\\
\rule{0pt}{3ex}
$(H^{*}(X), HF^{*}(\real{X},\real{X}))$  
&  & 
$(H^{*,*}(Y), H^{0,*}(Y))$  \\
\end{tabular}
\end{center}

If we restrict ourselves to  three point operations, we apparently obtain  
$\cH_c$ and $\widehat{\cH}_c$-module structure in respective sides
(which reminds Horava's version of quantum cohomology
of real varieties). 

For A-side of the mirror correspondence, the DG-operad structure that we 
have discussed  in previous section provides a promising candidate of 
extension this structure. However, the B-side of the story is more intriguing: 
Open-closed strings in B-model, their boundary conditions etc. are quite 
unclear to us.  S. Merkulov pointed out possible connections with Ran's work 
on relative obstructions, Lie atoms and their deformations \cite{r1,r2}

%
%

%

\begin{thebibliography}{99.}

\bibitem{br} G. Bredon, {\it Equivariant Cohomology Theories}. Lecture Notes 
in Mathematics 34, Springer Verlag, Berlin, 1967. 


\bibitem{c}   \"O. Ceyhan, {\it On moduli of pointed real curves of genus zero}.
{\em  Proceedings 13th Gokova Geometry Topology Conference} 2006, 1-38. 


\bibitem{th}   \"O. Ceyhan, {\it On moduli of pointed real curves of genus zero}.
Ph.D. Thesis. IRMA Preprints. Ref: 06019.


\bibitem{c1}  \"O. Ceyhan, {\it The graph homology of the moduli space of 
pointed real curves of genus zero}. to appear in Selecta Mathematica. 


\bibitem{c2}  \"O. Ceyhan, {\it On Solomon's reconstruction theorem for 
Gromov-Witten-Welshinger classes}. in preperation. 


\bibitem{cmirror}  \"O. Ceyhan, {\it `Realizing' mirror symmetry}. in preperation. 

\bibitem{DJS} M. Davis, T. Januszkiewicz, R. Scott, {\it Fundamental 
groups of blow-ups}.  Adv. Math.  177  (2003),  no. 1, 115--179.


\bibitem{dik}  A. Degtyarev, I. Itenberg, V. Kharlamov, {\it Real 
enriques surfaces}, Lecture Notes in Mathmetics 1746, Springer-Verlag, Berlin, 
2000. xvi+259 pp.


\bibitem{de}   S.L. Devadoss, {\it Tessellations of moduli spaces and the 
mosaic operad}. in `Homotopy invariant algebraic structures' (Baltimore, MD, 
1998),  91--114, Contemp. Math., 239, Amer. Math. Soc., Providence, RI, 1999.


\bibitem{dol}   I. Dolgachev, {\it Topics in Classical Algebraic Geometry. 
Part I}. lecture notes, available authors webpage 
{\it http://www.math.lsa.umich.edu/~idolga/topics1.pdf}


\bibitem{ehkr} P. Etingof,  A. Henriques, J. Kamnitzer, E. Rains, {\it 
The cohomology ring of the real locus of the moduli space of stable curves 
of genus 0 with marked points}. preprint math.AT/0507514.


\bibitem{fu}  K. Fukaya, Y.G. Oh,  H. Ohta,  K. Ono,  {\it Lagrangian 
intersection Floer theory: anomaly and obstruction}. preprint 2000. 



\bibitem{gh} P. Griffiths, J. Harris, {\it Principles of algebraic geometry}. Pure 
and Applied Mathematics. Wiley-Interscience [John Wiley and Sons], New York, (1978). 
xii+813 pp.


\bibitem{gm} A.B. Goncharov, Yu.I. Manin, {\it Multiple $\zeta$-motives and 
moduli spaces $\cmod{0,n}$}.  Compos. Math.  140 (2004), no. 1, 1--14.


\bibitem{hm} J. Harris, I. Morrison, {\it Moduli of curves}. Graduate Texts 
in Mathematics  vol 187, Springer-Verlag, NewYork, (1998), 366 pp.


\bibitem{hk} A. Henriques, J. Kamnitzer, {\it Crystals and coboundary categories}.
Duke Math. J. 132 (2006), no. 2, 191--216.


\bibitem{ho} P. Horava, {\it Equivariant topological sigma models}. 
Nuclear Phys. B 418 (1994), no. 3, 571--602.


\bibitem{iks} I. Itenberg, V. Kharlamov, E. Shustin, {\it A Caporaso-Harris 
type formula for Welschinger invariants of real toric Del Pezzo surfaces}. 
preprint math/0608549.


\bibitem{ks1} H. Kajiura, J. Stasheff, {\it Jim Homotopy algebras inspired by 
classical open-closed string field theory}. Comm. Math. Phys. 263 (2006), no. 3, 553--581.


\bibitem{ks2} H. Kajiura, J. Stasheff, {\it  Open-closed homotopy algebra in 
mathematical physics}. J. Math. Phys. 47 (2006), no. 2, 023506, 28 pp.


\bibitem{ks3} H. Kajiura, J. Stasheff, {\it Homotopy algebra of open-closed 
strings}. hep-th/ 0606283.


\bibitem{ka}  M. Kapranov, {\it The permutoassociahedron, MacLane's coherence 
theorem and asymptotic zones for KZ equation}.  J. Pure Appl. Algebra  85  (1993),  
no. 2, 119--142.


\bibitem{ke} S. Keel, {\it Intersection theory of moduli space of stable N-pointed 
curves of genus zero},  Trans. Amer. Math. Soc.  330  (1992),  no. 2, 545--574.


\bibitem{kon} M. Kontsevich, {it Homological algebra of mirror symmetry}. Proceedings 
of the International Congress of Mathematicians, Vol. 1, 2 (Z\"urich, 1994), 120--139, 
Birkh\"auser, Basel, 1995. 

\bibitem{km1} M. Kontsevich, Yu.- I.  Manin, {\it Gromov-Witten classes, quantum 
cohomology, and enumerative geometry}.  Comm. Math. Phys.  164  (1994),  no. 3, 
525--562.


\bibitem{km2} M. Kontsevich, Yu.-I.  Manin (with appendix by R. Kaufmann), 
{\it Quantum cohomology of a product}. Invent. Math. 124  (1996),  no. 1-3, 313--339.


\bibitem{knu} F.F. Knudsen, {\it The projectivity of the moduli space of stable curves. II. 
The stacks $M\sb{g,n}$}. Math. Scand.  52  (1983),  no. 2, 161--199.




\bibitem{liu} C.C.M. Liu, {\it Moduli J-holomorphic curves with Lagrangian boundary 
conditions and open Gromov-Witten invariants for an $S^{1}$-equivariant pair}. preprint
math.SG/0210257.


\bibitem{m}  Yu.-I. Manin, {\it Frobenius manifolds, quantum
cohomology and moduli spaces}. AMS Colloquium Publications
vol 47, Providence, RI, (1999), 303 pp.


\bibitem{m2} Yu.-I. Manin, {\it Gauge Field Theories and
Complex Geometry}. Grundlehren der Mathematischen Wissenschaften
289, Springer-Verlag, Berlin, (1997), 346 pp.


\bibitem{mer} S. A. Merkulov, {\it Operad of formal homogeneous spaces 
and Bernoulli numbers}. math/0708.0891.


\bibitem{mik1} G. Mikhalkin, {\it Counting curves via lattice paths in polygons}. 
C. R. Math. Acad. Sci. Paris 336 (2003), no. 8, 629--634.


\bibitem{mik2} G. Mikhalkin, {\it Enumerative tropical algebraic geometry in 
$\Bbb R\sp 2$}. J. Amer. Math. Soc. 18 (2005), no. 2, 313--377.


\bibitem{psw}  R. Pandharipande, J. Solomon, J. Walcher,
{\it Disk enumeration on the quintic 3-fold}. math.SG/0610901.


\bibitem{r1}  Z. Ran, {\it Lie Atoms and their deformation theory}. math/0412204.

\bibitem{r2}  Z. Ran, {\it Bernoulli numbers and deformations of schemes and maps}.
math/ 0609223.


\bibitem{s} J. Solomon,  {\it Intersection theory on the moduli space of 
holomorphic curves with Lagrangian boundary conditions}.  math.SG/0606429.


\bibitem{stalk}  J. Solomon, {\it A differential equation for the open Gromov-Witten potential}.
a talk given at {\em Real, tropical, and complex enumerative geometry workshop}
at CRM-Montreal.

\bibitem{vo} A. Voronov, {\it The Swiss-cheese operad}. Homotopy invariant algebraic structures (Baltimore, MD, 1998), 365--373, Contemp. Math., 239, Amer. Math. Soc., Providence, RI, 1999. 

\bibitem{w1} J.Y. Welschinger, {\it Invariants of real symplectic 4-manifolds and lower 
bounds in real enumerative geometry}. Invent. Math.  162  (2005),  no. 1, 195--234.


\bibitem{w2} J.Y. Welschinger, {\it Spinor states of real rational curves in real algebraic 
convex 3-manifolds and enumerative invariants}. Duke Math. J.  127  (2005),  no. 1, 89--121.

\bibitem{w3} J.Y. Welschinger,  {\it Invariants of real symplectic four-manifolds out 
of reducible and cuspidal curves}. Bull. Soc. Math. France 134 (2006), no. 2, 287--325.

\bibitem{w4} J.Y. Welschinger,  {\it Towards relative invariants of real symplectic 
four-manifolds.} Geom. Funct. Anal. 16 (2006), no. 5, 1157--1182.
\end{thebibliography}
%


\end{document}